\documentclass{article}

\usepackage[english]{babel}

\usepackage[letterpaper,top=2cm,bottom=2cm,left=3cm,right=3cm,marginparwidth=1.75cm]{geometry}

\usepackage{amsmath}
\usepackage{graphicx}
\usepackage[colorlinks=true, allcolors=blue]{hyperref}
\usepackage{multirow}
\usepackage{booktabs}
\usepackage{caption}
\usepackage{authblk}
\usepackage{icomma}

\title{Tree statistics improve neural simulation-based inference of epidemiological parameters over case counts alone}
\author[1]{Zena Lapp}
\author[1]{Emma E.~Goldberg}
\author[2]{John R. Tipton}
\author[2]{Thea Sukianto}
\author[2]{Dave Osthus}
\author[3]{Lauren A.~Castro}
\affil[1]{Theoretical Biology and Biophysics Group, Los Alamos National Laboratory, Los Alamos, New Mexico, United States of America}
\affil[2]{Statistics Group, Los Alamos National Laboratory, Los Alamos, New Mexico, United States of America}
\affil[3]{Computational Intelligence and Modeling Group, Los Alamos National Laboratory, Los Alamos, New Mexico, United States of America}

\begin{document}
\maketitle

\begin{abstract}
\noindent During an infectious disease outbreak, accurate estimates of epidemiological parameters such as the reproduction number and infection duration can inform the public health response. Recently, neural simulation-based inference (NSBI) methods have become popular for estimating these parameters due to lower computational costs and more input data flexibility than traditional methods. Input data modalities that have been evaluated independently in an NSBI framework include case counts and transmission trees. Here, we evaluate whether combining information from these two data types outperforms using information from each individually with NSBI. To do so, we simulated stochastic susceptible-infectious-recovered (SIR) outbreaks across a range of basic reproduction numbers ($R_0$) and average infection durations ($D$) to generate training, validation, and testing datasets. We then calculated, for each simulated outbreak, summary statistics from case count trajectories and from transmission trees. Using these features, individually and in combination, we performed a type of NSBI called neural ratio estimation (NRE) to estimate $R_0$ and $D$. We also calculated maximum likelihood estimates of the parameters by fitting a deterministic SIR model using the case count trajectory. All three NRE models outperformed the baseline method. Furthermore, incorporating tree summary statistics improved the estimation of both $R_0$ and $D$ over case count summaries alone. This suggests that trees provide valuable additional information for estimating epidemiological parameters above and beyond case counts alone.
\end{abstract}



\section{Author Summary}
During an infectious disease outbreak, understanding how readily an infection spreads and how long people remain infectious is critical in guiding the public health response. Case counts provide valuable information about these characteristics, but they do not describe who infected whom. We investigated whether transmission trees, which capture these connections, provide useful information beyond case counts alone. We simulated outbreaks with different levels of disease spread and average times to recovery, then trained machine learning models to estimate these characteristics using summaries of case counts, transmission trees, or both with uncertainty quantification of these estimates. Because we knew the values used to generate each outbreak, we could directly assess estimation accuracy. All three machine learning models outperformed a conventional approach based on case counts. Among the machine learning models, including transmission tree information improved estimates of how readily the disease spreads and the average time to recovery. Our findings show that connections between infections can reveal information that case counts alone do not fully capture.

\clearpage
\newgeometry{top=0.85in,left=1in,right=1in,footskip=0.75in}

\section{Introduction}
Accurate estimation of epidemiological parameters is essential for an informed public health response. Estimates of infection duration can inform contact tracing and the timing or duration of interventions \cite{cuartero_impact_2026}, while reproduction numbers characterize the potential for disease spread. Together, these parameters help researchers forecast case counts and assess the potential effects of interventions. However, estimating them from noisy observations can be difficult, and inaccurate estimates can result in ineffective public health decisions \cite{ridenhour_unraveling_2014}.

Epidemiological parameter estimation commonly involves fitting compartmental models, such as susceptible-infectious-recovered (SIR) models, to case count or hospitalization time series using maximum likelihood or Bayesian methods \cite{boonpatcharanon_estimating_2022}. These approaches are straightforward when the likelihood can be evaluated, but for stochastic epidemic models the likelihood can be difficult to compute, particularly when infection and recovery events are only partially observed. Particle Markov chain Monte Carlo (MCMC) addresses this challenge by using particle filtering to estimate the likelihood within Bayesian inference, although repeated filtering throughout the MCMC procedure can be computationally demanding. Approximate Bayesian computation (ABC) \cite{beaumont2002approximate} provides a likelihood-free alternative by using simulations to bypass likelihood evaluation altogether. ABC, however, can still require large numbers of simulations and typically requires choosing summary statistics, a discrepancy measure, and an acceptance tolerance, all of which can be challenging to defensibly select. Neural simulation-based inference (NSBI) \cite{cranmer_frontier_2020, zammit-mangion_neural_2025} also uses simulations to enable inference without explicit likelihood evaluation, but instead trains neural networks to learn an inferential relationship between model parameters and simulated data. This avoids the need for an explicit discrepancy measure or acceptance tolerance and, in the amortized setting, shifts much of the computational cost to an initial simulation and training stage. Once trained, the network can be reused to perform rapid inference for new datasets without retraining for each analysis, as long as the mechanisms and observation conditions of the new outbreak are represented in the training data \cite{cranmer_frontier_2020, radev_outbreakflow_2021}. This separation creates opportunities to prepare inference tools in advance of an outbreak and apply them repeatedly as surveillance data become available. It also makes extensive simulation-based evaluation practical, allowing estimation accuracy and uncertainty calibration to be assessed across many outbreak scenarios. 

Another benefit of NSBI is it can accommodate a wide range of input representations, making it appealing for parameter estimation from diverse sources of information. Recent studies have applied NSBI methods to infer epidemiological parameters from outbreak case counts \cite{jang_comparative_2026, radev_outbreakflow_2021, chatha_neural_2026}, pathogen sequence alignments \cite{garot2026teddy}, and phylogenetic trees \cite{zarebski_amortized_2026, pinotti2026simulation, perez_phylocnn_2026}. The success of inference from these individual data sources raises the question of what can be gained by combining them. Case counts describe the progression of an outbreak, while trees capture relationships among infections that aggregate counts do not retain. Joint inference from epidemiological and phylogenetic information has already been explored through particle MCMC approaches, including the framework of Rasmussen et al. \cite{rasmussen2011inference}, EpiInf \cite{vaughan2019estimating}, and EpiFusion \cite{judge_epifusion_2024}, as well as through the likelihood approximation used by TimTam \cite{zarebski2022computationally}. EpiFusion explicitly compares case incidence alone, phylogenetic trees alone, and their combination, demonstrating benefits of integrating the two sources. These studies motivate a corresponding comparison within NSBI, where a common inference framework can accommodate information from either source or both while retaining the advantage of rapid inference after training.

Here, we systematically compare case count summaries, transmission tree summaries, and their combination for jointly estimating the basic reproduction number ($R_0$) and average infection duration ($D$) with NSBI. 
We used neural ratio estimation (NRE) to estimate and quantify uncertainty in both parameters. Using simulated stochastic SIR outbreaks with known parameter values, we evaluated estimation accuracy and uncertainty calibration and compare the NSBI models with a baseline maximum likelihood approach that fits a deterministic SIR model to case counts. This design allows us to assess two distinct questions: (1) whether NSBI improves inference from case counts relative to the baseline, and (2) how the inclusion of transmission tree summaries changes inference relative to case count summaries alone in an NSBI framework.

\section{Results}
We first describe the simulated outbreaks (Sections \ref{subsec:simdata}) and examine associations between epidemiological parameters and summary statistics derived from case counts and transmission trees (Section \ref{subsec:simcor}). 
We then compare case count-based NSBI with the maximum likelihood baseline (Section \ref{subsec:baselineresults}) before assessing the additional value of transmission tree summaries within NSBI (Section \ref{subsec:treeresults}). 
Finally, we assess how performance varied across the parameter space to identify where transmission tree information provides the greatest benefit (Section \ref{subsec:results_details}).

\subsection{Simulated data provide case count trajectories and transmission trees}
\label{subsec:simdata}

\begin{figure}
    \centering
    \includegraphics[width=1\linewidth]{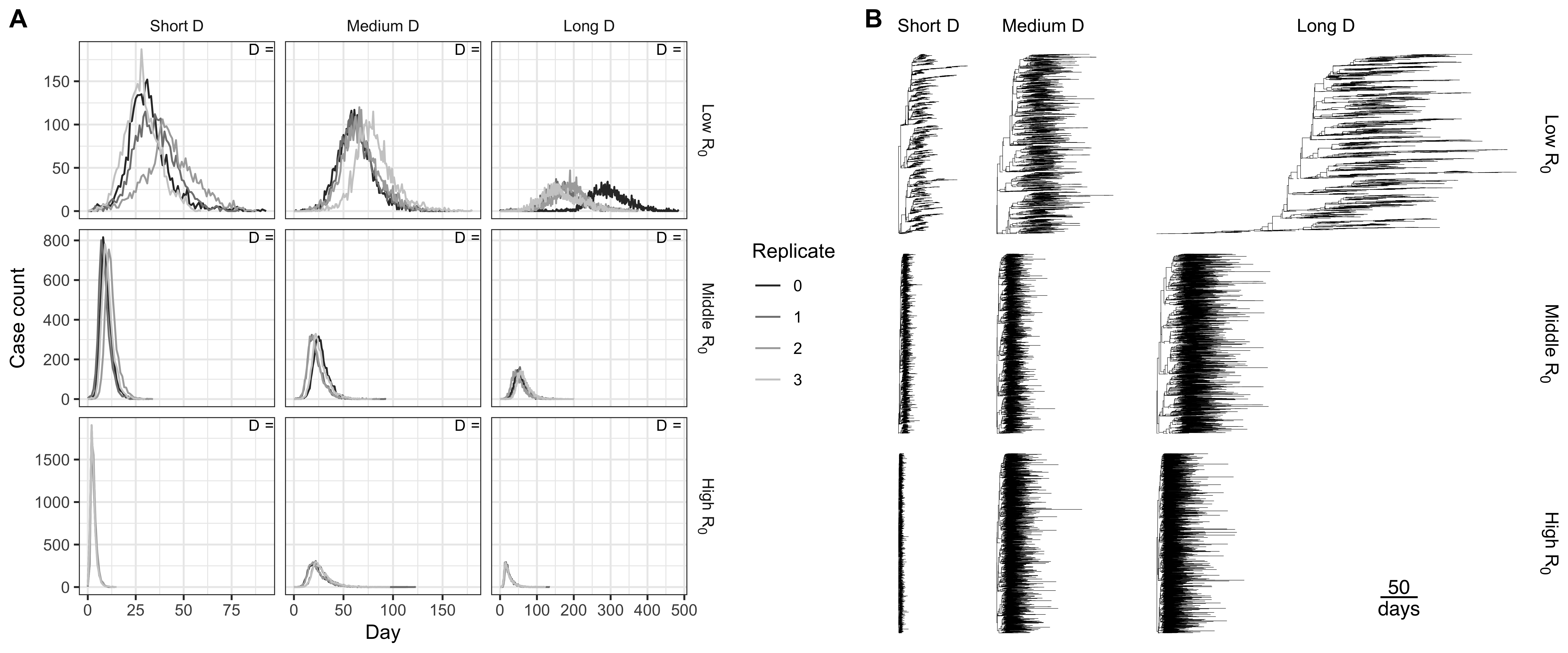}
    \caption{Examples of simulated test data. (A) Daily case count trajectories. (B) Corresponding transmission trees for replicate 0 in panel A.}
    \label{fig:simulation_examples}
\end{figure}

We used Latin hypercube sampling (LHS) \cite{mckay_comparison_1979} of the basic reproduction number, $R_0$, and average infection duration, $D$, to generate $25,000$ training parameter sets and $5,000$ testing parameter sets spanning the two-dimensional parameter space. For each parameter combination, we simulated 4 stochastic SIR outbreaks using ReMASTER \cite{vaughan2024} in BEAST2 \cite{bouckaert2019} with complete sampling of individuals upon moving to the recovered compartment. This resulted in $100,000$ training simulations and $20,000$ testing simulations. Each simulation yielded a daily case count trajectory and a corresponding transmission tree. These paired data allowed us to compare inference from the two information sources, individually and together, for the same outbreaks. Example trajectories and their corresponding trees illustrate the variation across parameter settings and stochastic replicates (Fig \ref{fig:simulation_examples}). Outbreaks with lower $R_0$ values tended to produce noisier case count trajectories, exhibit more replicate-to-replicate variability, and infect a smaller proportion of the susceptible population, while outbreaks with longer average infection durations tended to last longer (Fig \ref{fig:simulation_examples}A).
The corresponding transmission trees also reflect these differences in outbreak dynamics (Fig \ref{fig:simulation_examples}B). Furthermore, higher $R_0$ values produced more temporally concentrated branching near the root, consistent with rapid early transmission.

\subsection{More tree summaries than case count summaries correlate strongly with epidemiological parameters}
\label{subsec:simcor}

\begin{figure}[h!]
    \centering
    \includegraphics[width=0.75\linewidth]{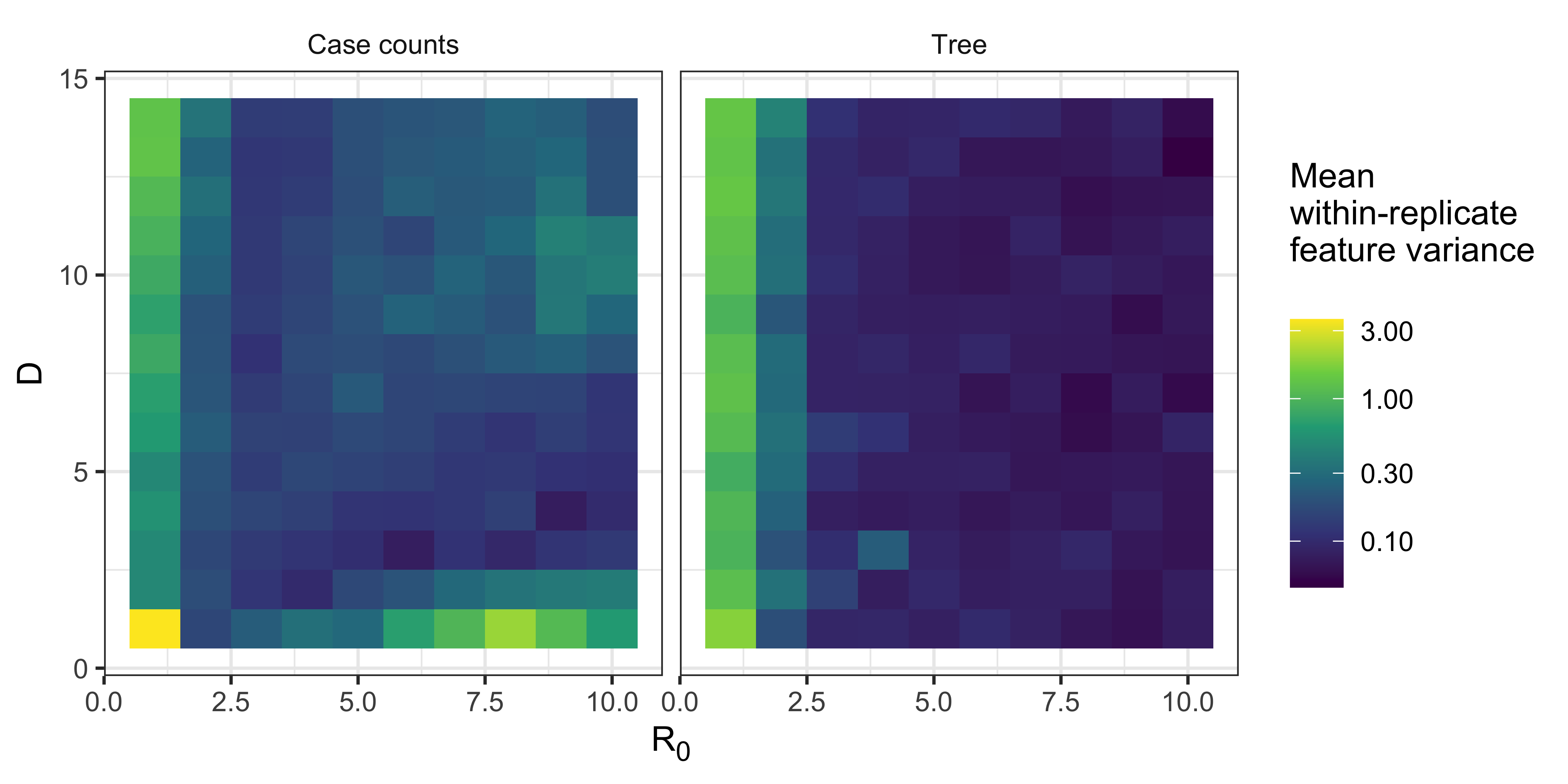}
    \caption{Mean within-replicate feature variance by parameter for case count features and tree features. Parameters were rounded to the nearest whole number. Variance was computed per-simulation, per-feature on the z-score normalized feature values.}
    \label{fig:feature_variance}
\end{figure}

\begin{figure}
    \centering
    \includegraphics[width=1\linewidth]{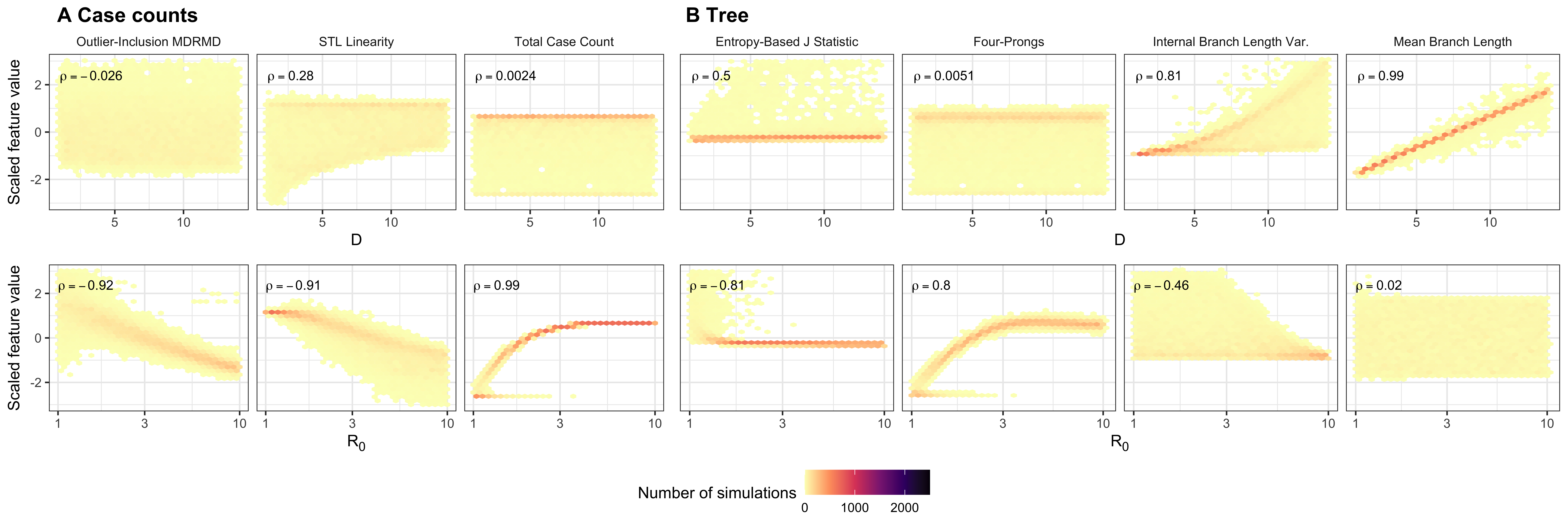}
    \caption{Correlation between parameter values and example features for test data. (A) Case count features. This includes all features that have a Spearman correlation of $|\rho|\geq 0.8$ with at least one parameter. Note that in our case the total case count feature is equivalent to the number of lineages tree feature. (B) Tree features. These features were selected for visualization by filtering to those that had $|\rho|\geq 0.8$ with at least one parameter, and then selecting one representative feature for each set of features that were correlated by $|\rho|\geq 0.8$. For visualization purposes, the scaled feature values were filtered to be within 3 standard deviations of the mean. However, the Spearman correlation coefficients were computed on all values. All case count summaries and tree summaries, with their corresponding correlations, are presented in Tables \ref{tab:tsfeatures} and \ref{tab:treefeatures}, respectively.}
    \label{fig:correlations}
\end{figure}

For each simulated outbreak, we calculated 49 summary statistics from the daily case count time series and 51 from the transmission tree (Tables \ref{tab:tsfeatures} and \ref{tab:treefeatures}). First, to investigate replicate variability, we calculated the variance of each z-score normalized summary statistic for each set of replicates. The variance tended to be low (median=0.016, mean=0.321) although the distribution was heavily skewed (max=4895). Furthermore, variance tended to be higher at lower $R_0$ (Fig \ref{fig:feature_variance}), consistent with what we observed in Fig \ref{fig:simulation_examples}. Next, to explore associations between summary statistics and epidemiological parameters, we calculated Spearman correlations ($-1 \leq \rho \leq 1$) between each summary statistic and $R_0$ or $D$, and defined strong correlation as those with $|\rho|\geq 0.8$.
Summaries can be highly correlated with one parameter but not another (Fig \ref{fig:correlations}). For instance, the mean branch length of the transmission tree is highly correlated with $D$ ($\rho = 0.99$) but essentially uncorrelated with $R_0$ ($\rho = 0.02$) (Fig \ref{fig:correlations}B).
For both parameters, a larger proportion of tree summaries than case count summaries had a strong correlation. For $R_0$, strong correlations were observed for $20/51$ ($39.2\%$) tree summaries compared with $3/49$ ($6.1\%$) case count summaries. For $D$, $5/51$ ($9.8\%$) tree summaries, all branch length statistics, met the threshold, whereas none of the case count summaries did. These correlation patterns suggest that including tree summaries could improve estimation over case count summaries alone, motivating the model comparisons that follow.

\subsection{Case count-based NSBI improves on the baseline method}
\label{subsec:baselineresults}

We first compared an NSBI model using case count summaries alone to a baseline maximum likelihood approach that fit a deterministic SIR model to the case count time series. We evaluated point estimate accuracy using absolute error, uncertainty using interval width, interval calibration using empirical coverage of nominal 95\% intervals, and overall interval performance using the interval score (details in Section \ref{sec:methods}). The NSBI model provided marginal highest posterior density (HPD) intervals, while the baseline provided confidence intervals.

\begin{figure}
    \centering
    \includegraphics[width=1\linewidth]{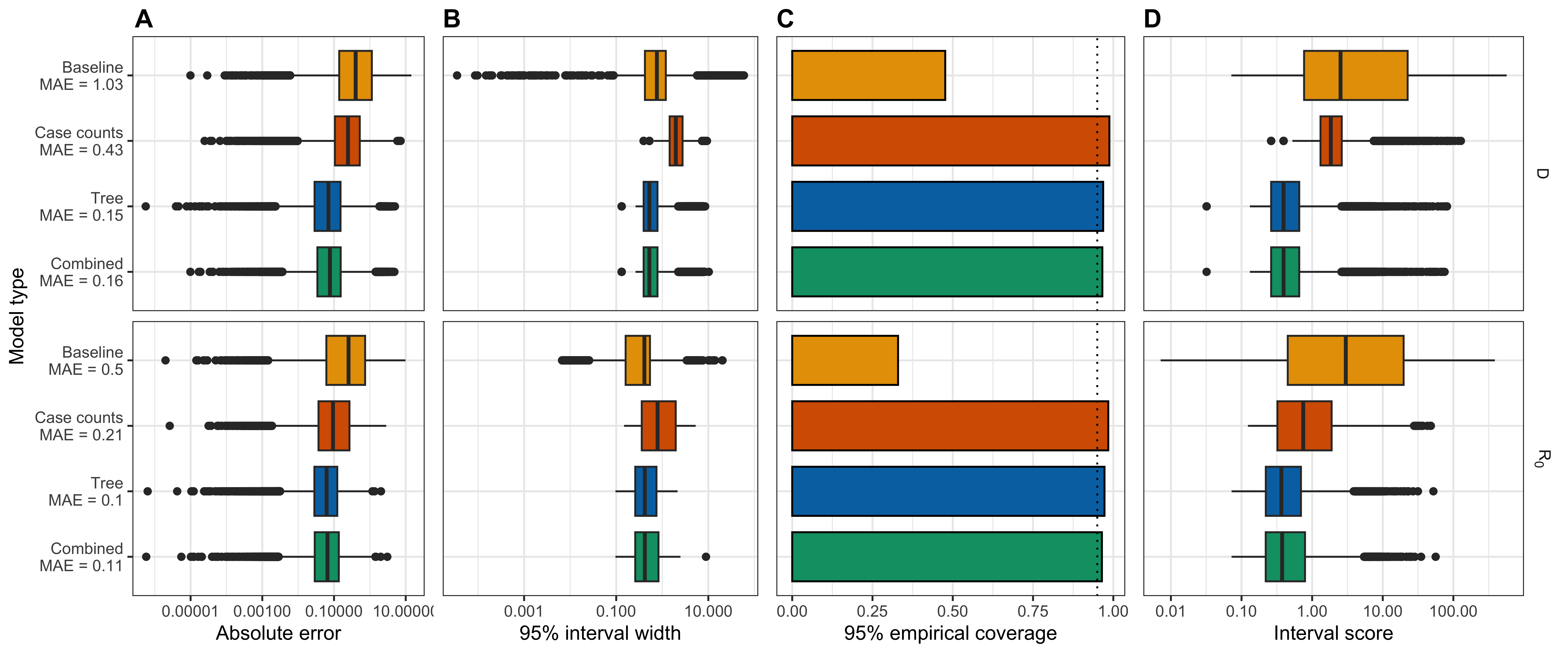}
    \caption{Model evaluation on the test set. (A) Absolute error of true value compared to point estimate. 
    The mean absolute error (MAE) is shown in the plot's label. (B) $95\%$ marginal highest posterior density (HPD) interval width for neural network models, $95\%$ confidence interval for baseline. (C) $95\%$ marginal empirical coverage. (D) Interval score (smaller is better). $D$ is infection duration and $R_0$ is basic reproduction number. The baseline estimate uses the maximum likelihood estimate (MLE) of a deterministic SIR model fit to the case counts time series. All other models use NSBI, with either case count summaries only, tree summaries only, or both case count and tree summaries combined as the available information set. We see that, for almost every metric and both parameters, the NSBI models perform better than the baseline, and the NSBI models that use tree-based information outperform the NSBI model that only uses case count information.}
    \label{fig:evaluation}
\end{figure}

For both parameters, $R_0$ and $D$, case count NSBI produced more accurate point estimates ($MAE_D=0.43$, $MAE_{R_0}=0.21$) than the baseline ($MAE_D=1.03$, $MAE_{R_0}=0.50$) (Fig \ref{fig:evaluation}A), with case count NSBI having a lower absolute error than the baseline in $67\%$ of test samples for $D$ and $72\%$ of test samples for $R_0$. While the baseline produced narrower uncertainty intervals than NSBI (Fig \ref{fig:evaluation}B), they were poorly calibrated and exhibited severe undercoverage (coverage $<50\%$) (Fig \ref{fig:evaluation}C). Although case count NSBI produced wider uncertainty intervals than the baseline, its intervals were much better calibrated (coverage slightly greater than $95\%$). The baseline's combination of lower accuracy and overly narrow uncertainty intervals resulted in worse interval scores than those of case count NSBI (Fig \ref{fig:evaluation}D). These results establish that NSBI has higher accuracy and interval calibration than the baseline even when inference relies solely on case counts.

\subsection{Tree summaries improve parameter inference, while case count summaries add little beyond trees}
\label{subsec:treeresults}

Having established the performance of NSBI using case counts alone, we next assessed whether transmission tree summaries provide additional benefits. We compared three NSBI models using case count summaries, transmission tree summaries, or both to jointly estimate $R_0$ and $D$. All three models were evaluated using the same metrics and marginal HPD interval construction described above. For both $R_0$ and $D$, the NSBI model using tree summaries produced more accurate point estimates ($MAE_D=0.16$, $MAE_{R_0}=0.10$) and narrower intervals than NSBI using case count summaries alone (Fig \ref{fig:evaluation}A,B), with the tree model having a lower absolute error in $82\%$ of test samples for $D$ and $65\%$ of test samples for $R_0$. This trend held when comparing the case count model to the model that combined case count and tree summaries. Furthermore, all three NSBI models achieved coverage slightly above the nominal 95\% level (Fig \ref{fig:evaluation}C), indicating that the narrower intervals obtained with tree summaries did not come at the expense of undercoverage. Models incorporating tree summaries also achieved lower interval scores (Fig \ref{fig:evaluation}D), demonstrating a better balance between accuracy, interval width, and coverage. Overall, NSBI methods that used tree summary information outperformed NSBI methods that did not. 

Combining case count and tree summaries provided no appreciable improvement over using tree summaries alone, with tree features only slightly out-performing the combined set (Fig \ref{fig:evaluation}; MAE difference of 0.01). This finding is consistent with the structure of our simulated data. The case count time series is itself a summary of the full transmission tree. Although the selected tree summaries need not retain all information in that time series, the comparison suggests that case count summaries added little information that the NSBI models could exploit beyond the selected tree summaries in this setting. However, in settings where the full transmission tree is not observed and therefore the case count time series is not a deterministic summary of the transmission tree, this finding may not hold (see Section \ref{sec:discussion}).

\subsection{Tree features improve inference at higher values of $R_0$ and $D$}
\label{subsec:results_details}
\begin{figure}[h!]
    \centering
    \includegraphics[width=1\linewidth]{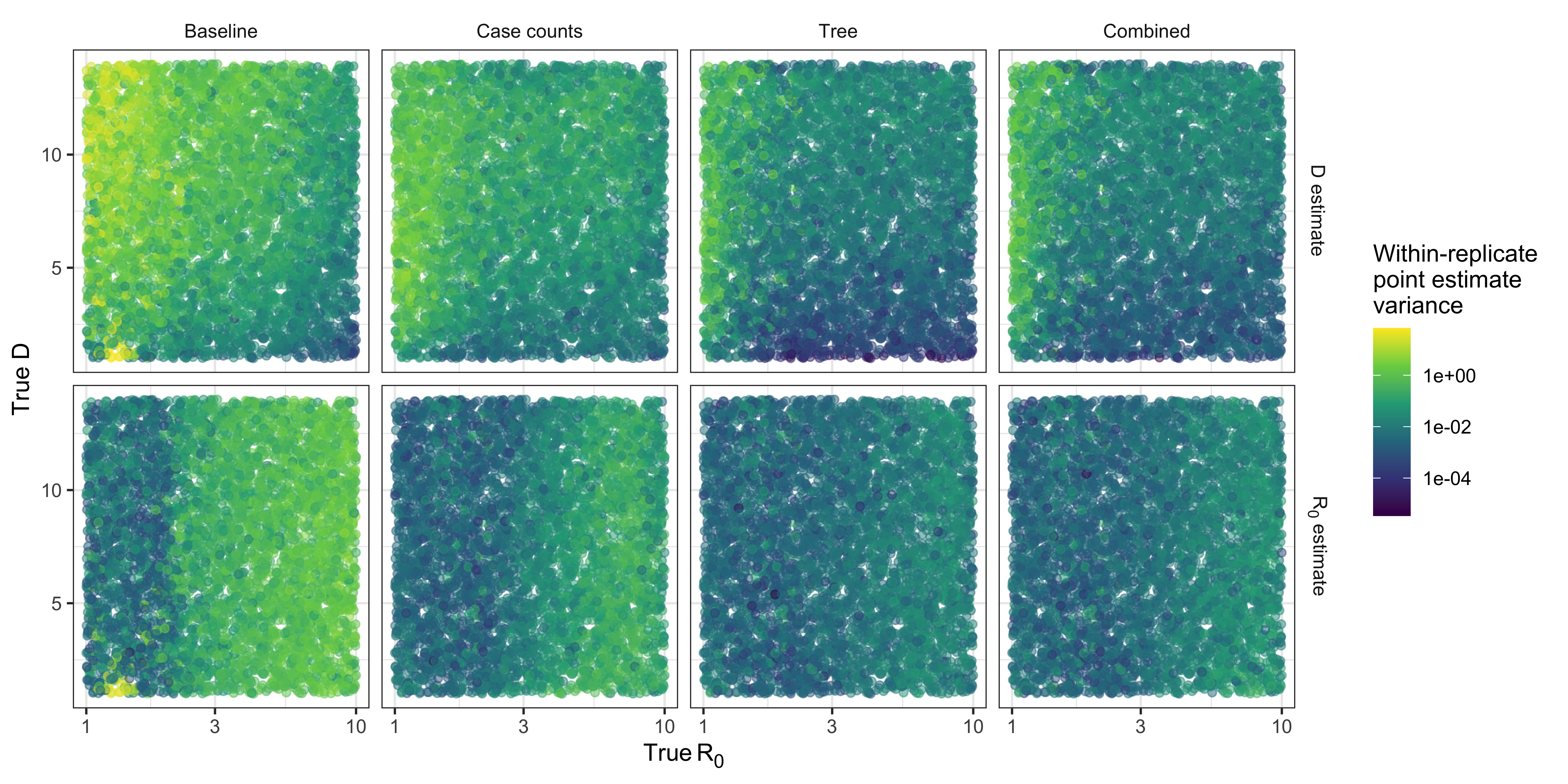}
    \caption{Within-replicate point estimate variance across parameter values.}
    \label{fig:estimate_variance}
\end{figure}

\begin{figure}
    \centering
    \includegraphics[width=1\linewidth]{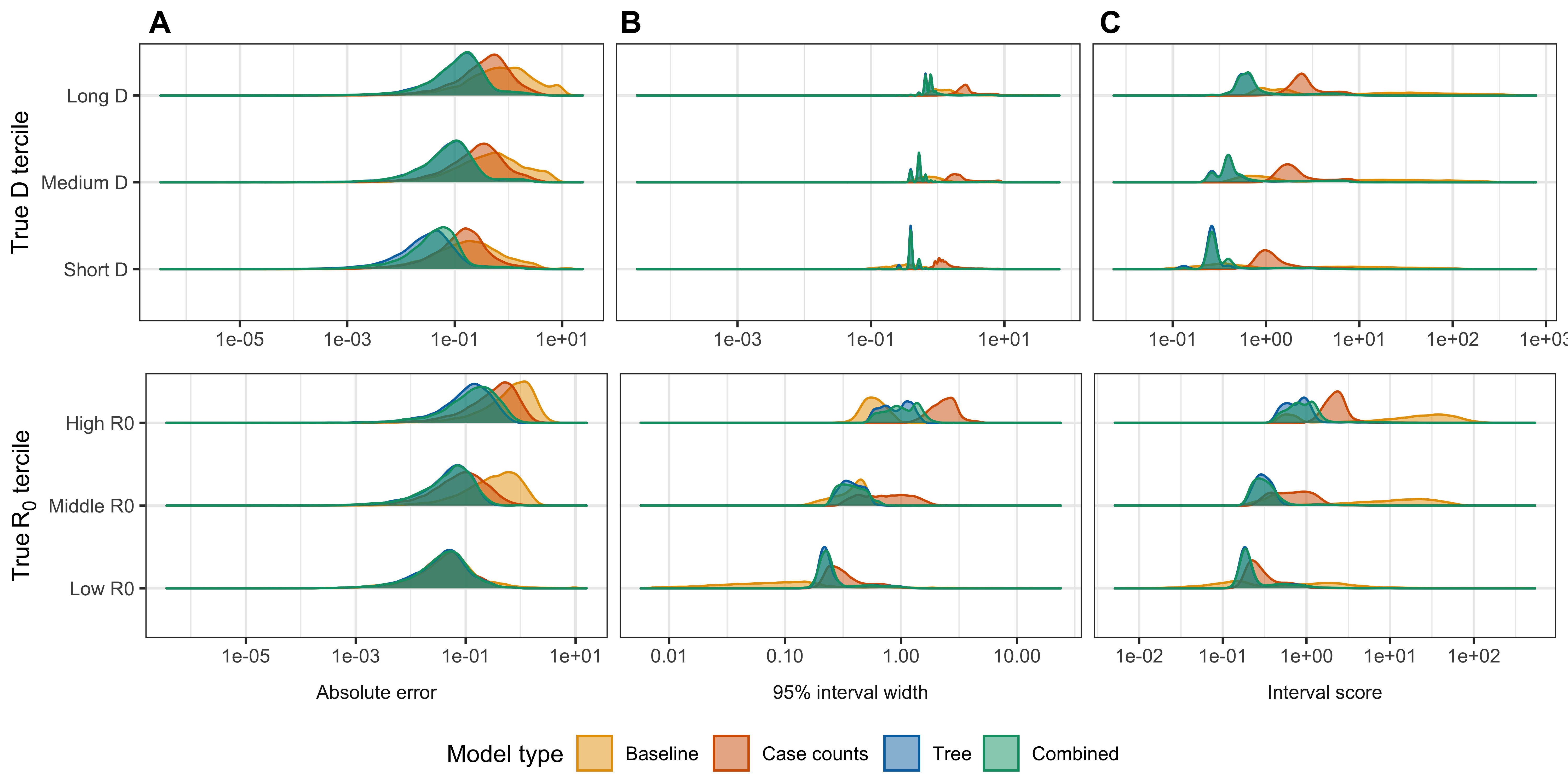}
    \caption{Evaluation metrics (columns) as a function of parameter value (y-axis) for $D$ (top) and $R_0$ (bottom) for the different models (colors).}
    \label{fig:eval_across_params}
\end{figure}

We next investigated how well the inference methods performed as a function of $R_0$ and $D$. First, within a set of replicates, we observed higher variance in the $D$ point estimates with lower true values of $R_0$ and higher true values of $D$, and higher variance in the $R_0$ point estimates with higher true values of $R_0$ (Fig \ref{fig:estimate_variance}). Second, across all evaluation metrics and inference methods, we observed a decrease in performance with increasing value for both parameters, be that accuracy (Fig \ref{fig:eval_across_params}A) interval width (Fig \ref{fig:eval_across_params}B), or interval score (Fig \ref{fig:eval_across_params}C). These observations align with the observed parameter-feature correlations (Fig \ref{fig:correlations}). For higher values of $D$, there is more uncertainty in the features (e.g. mean branch length). For higher values of $R_0$, feature values tend to be similar, and thus less informative across a range of $R_0$ values (e.g. total case count). Finally, at small values of true $R_0$, all four inference methods had similar absolute errors (Fig \ref{fig:eval_across_params}A), while at larger values the models that included tree features performed better, and the case count model performed better than the baseline. For $D$, on the other hand, models that included tree features had lower absolute errors across the entire range of $D$, while the case count and baseline models had higher absolute errors.

\section{Discussion}
\label{sec:discussion}

Our simulation study demonstrates the value of transmission tree summaries for estimating the basic reproduction number, $R_0$, and average infection duration, $D$, within an NSBI framework. All NSBI models improved on the deterministic SIR maximum likelihood baseline, with models incorporating tree summaries producing more accurate estimates and narrower uncertainty intervals than the NSBI model using case count summaries alone, while maintaining coverage near the nominal level. Combining case count and tree summaries, however, offered little additional improvement over using tree summaries alone. Together, these findings highlight the usefulness of NSBI, and of tree statistics, for epidemiological parameter inference.


The improvements of tree summaries over case counts, and of  NSBI over baseline methods, is consistent with previous observations. First, we observed an improvement in inference when using tree summaries in an NSBI framework. This is consistent with previous work demonstrating benefits from combining epidemiological and phylogenetic information using particle MCMC \cite{judge_epifusion_2024}. 
Second, we observed improved inference of NSBI over a deterministic baseline model. While this comparison does not establish superiority of our NSBI method over other inference methods that account for outbreak stochasticity, other studies have shown that NSBI performs better than more computationally intensive baseline methods. It performs better than ABC given a fixed computational budget \cite{jang_comparative_2026}, produces more accurate estimates with fewer simulations than ABC \cite{chatha_neural_2026}, and results in less biased estimates than a BEAST2 analysis \cite{zarebski_amortized_2026}. Taken together, these results highlight the competitiveness of NSBI relative to other standard methods in terms of both accuracy and computational efficiency. 


Our study was designed as a proof of concept for integrating case count and transmission tree information within an NSBI framework, and several aspects of its design limit direct application to public health surveillance. We evaluated inference using completely sampled case count trajectories and transmission trees simulated under an SIR model. 
Having a entirely complete and accurate transmission tree provides less noisy information about $R_0$ than would be the case in empirical data, where a subset of infections would be sampled, and even fewer sequenced. In addition, individuals were sampled onto the transmission tree at the time of recovery, providing relatively direct information about $D$ and likely making it easier to estimate than under more realistic sampling processes. Also, the similar performance of the tree-only and combined models here reflects the fact that the case count time series can be derived from the tip dates of the full transmission tree, resulting in overlapping information between the two data sources. In practice, case surveillance usually includes infections absent from the tree, and the phylogenetic tree itself represents an imperfectly reconstructed representation of transmission. The relative value of the two sources may therefore differ substantially under more realistic observation processes. Real-world analyses must contend with differences between transmission and phylogenetic trees, uncertainty in tree reconstruction, and delays and biases in both case reporting and sequencing. Finally, to perform NSBI on pathogens that spread via mechanisms other than SIR dynamics would require a model trained on simulations that incorporate those different mechanisms. These limitations motivate evaluation of NSBI under more realistic epidemic and observation processes.

The flexibility of NSBI inputs means that other data representations or data modalities can be used for inference instead of, or in addition to, the representations that we include here. We used predefined time series and tree summaries to compare the information sources, but neural architectures can instead learn representations from time series or tree structures directly \cite{radev_outbreakflow_2021, zarebski_amortized_2026, perez_phylocnn_2026}. Such representations could retain information lost through the selected summaries. A broader framework could also integrate case, hospitalization, and death time series with seroprevalence measurements, vaccination histories, and transmission or phylogenetic trees. These extensions would require suitable architectures and training data, but would not require analytically tractable likelihoods. Realizing this flexibility would require forward models that coherently connect the available information sources, representing an exciting direction for future work.

In summary, our findings support NSBI as a practical framework for evaluating and integrating outbreak information, with transmission tree summaries providing clear benefits in the setting studied here. Future work should assess how these benefits change with incomplete observations, reconstructed trees, and more realistic epidemic dynamics, and whether learned representations and additional surveillance streams improve inference further.

\section{Methods}
\label{sec:methods}

We first describe the epidemic model and simulation design (Section \ref{subsec:simulation_design}), followed by the summary statistics and preprocessing steps used to construct model inputs (Section \ref{subsec:preprocessing}). We then present the neural ratio estimation framework (Section \ref{subsec:nre}) and maximum likelihood baseline (Section \ref{subsec:baseline_methods}). Finally, we describe the evaluation criteria (Section \ref{subsec:evaluation}) and software and resources needed to reproduce the analyses (Section \ref{subsec:software}).

\subsection{Epidemic model and simulation design}
\label{subsec:simulation_design}

We simulated stochastic SIR outbreaks to evaluate joint estimation of the basic reproduction number, $R_0$, and average infection duration, $D$. Simulations were performed using ReMASTER \cite{vaughan2024} in BEAST2 \cite{bouckaert2019}. The model described a closed population of susceptible ($S$), infectious ($I$), and recovered ($R$) individuals, with constant population size $N=S+I+R$. Under frequency-dependent transmission, infection and recovery events occurred according to
\begin{align}
    (S,I,R) &\longrightarrow (S-1,I+1,R)
    &&\text{at rate } \frac{\beta SI}{N},\\
    (S,I,R) &\longrightarrow (S,I-1,R+1)
    &&\text{at rate } \gamma I,
\end{align}
where $\beta>0$ is the transmission rate parameter and $\gamma>0$ is the recovery rate. Under this convention, the basic reproduction number in a fully susceptible population and the average infection duration were
\begin{equation}
    R_0=\frac{\beta}{\gamma},
    \qquad
    D=\frac{1}{\gamma}.
\end{equation}
We obtained the simulation rates from the target parameters using $\gamma=1/D$ and $\beta=R_0/D$.

We used Latin hypercube sampling (LHS) \cite{mckay_comparison_1979} with the maximin criterion \cite{johnson_minimax_1990} to obtain space-filling coverage of the parameter space, implemented using the Python package \texttt{pyDOE2} \cite{noauthor_clicumupydoe2_2026}. Sampling was performed in the coordinates $(\log_{10}R_0,D)$ over
\begin{equation}
    (\log_{10}R_0,D)\in(0,1)\times(1,14),
\end{equation}
corresponding to $R_0$ between 1 and 10 and $D$ between 1 and 14 days.

We generated 25,000 parameter sets for model development and a separate additional 5,000 parameter sets for testing. Four stochastic replicates per parameter set yielded 100,000 simulations for development and 20,000 simulations for testing. The 100,000 development runs were divided into training (90\%) and validation (10\%) subsets, with all replicates of each parameter set assigned to the same subset.

Each simulation began with $S(0)=4,999$, $I(0)=1$, and $R(0)=0$, giving $N=5,000$, and terminated when no infectious individuals remained. Every infected individual was sampled upon recovery, yielding paired epidemic trajectories and transmission trees with complete sampling. Outbreaks that ended with fewer than ten recovered individuals were discarded and simulated again.

\subsection{Summary statistics and preprocessing}
\label{subsec:preprocessing}

Simulated outbreaks varied in duration and number of sampled individuals. To obtain fixed-length inputs without requiring alignment of trajectories or trees, we represented each outbreak using case count summaries, transmission tree summaries, or their combination. To limit computational cost, we retained only summaries that required less than 0.1 seconds to compute for a 1,000-day outbreak (case count summaries) or a 10,000-person outbreak (tree summaries).

Case count time series were constructed by aggregating recovery events into one-day bins. We extracted time series summaries using the R package \texttt{tsfeatures} \cite{tsfeatures}. We also included the total number of sampled individuals, which is obtained by summing the case counts, to ensure that this quantity was available to both case count-based and tree-based models. The resulting set contained 49 case count summaries (Table \ref{tab:tsfeatures}). From each transmission tree, we extracted 51 topology and branch length summaries using the R packages \texttt{ape} \cite{paradis_ape_2019} to load the trees and \texttt{treestats} \cite{treestats} to compute the statistics (Table \ref{tab:treefeatures}). The combined feature set contained summaries from both sources.

Some summaries were undefined for certain outbreaks, such as those with very short durations. To retain these outbreaks, we constructed a binary mask for each feature vector, assigning one to finite feature values and zero to missing or non-finite values. Each feature was standardized by subtracting its training set mean and dividing by its training set standard deviation, calculated using only valid entries. The same normalization statistics were applied to the validation and test sets. Missing or non-finite entries were represented by zero in the standardized feature vector, which was then concatenated with its binary mask. This allowed the network to distinguish missing entries from valid values near the training set mean.

The parameters supplied to the neural network classifier were also standardized using training set means and standard deviations. Infection duration, $D$, was standardized on its original scale, while $R_0$ was first transformed to $\log_{10}R_0$ to match the coordinates used for LHS. The same transformations were used during validation, testing, and posterior evaluation.

\begin{table}[h!]

\caption{\label{tab:tsfeatures}Spearman correlation of case count features with parameters}
\centering
\begin{tabular}[t]{llrr}
\toprule
Feature & \texttt{tsfeatures} fn & $R_0$ & $D$\\
\midrule
\addlinespace[0.3em]
\multicolumn{4}{l}{\textbf{Autocorrelation \& Stationarity}}\\
\hspace{1em}ACF (Lag 1) & acf\_features() & 0.42 & 0.42\\
\hspace{1em}ACF (Lags 1-10) & acf\_features() & -0.17 & 0.59\\
\hspace{1em}ACF (1st Diff, Lag 1) & acf\_features() & 0.71 & -0.50\\
\hspace{1em}ACF (1st Diff, Lags 1-10) & acf\_features() & 0.40 & -0.35\\
\hspace{1em}ACF (2nd Diff, Lag 1) & acf\_features() & 0.45 & -0.51\\
\hspace{1em}ACF (2nd Diff, Lags 1-10) & acf\_features() & -0.25 & 0.30\\
\hspace{1em}PACF (Lags 1-5) & pacf\_features() & 0.60 & -0.16\\
\hspace{1em}PACF (1st Diff, Lags 1-5) & pacf\_features() & 0.18 & -0.37\\
\hspace{1em}PACF (2nd Diff, Lags 1-5) & pacf\_features() & -0.66 & 0.48\\
\hspace{1em}First Minimum of ACF & autocorr\_features() & 0.51 & 0.35\\
\hspace{1em}First Zero of ACF & firstzero\_ac() & -0.24 & 0.69\\
\hspace{1em}KPSS Unit Root Test & unitroot\_kpss() & 0.04 & 0.75\\
\hspace{1em}Phillips-Perron Test & unitroot\_pp() & 0.05 & 0.11\\
\hspace{1em}Autocorrelation (Lag 9) & autocorr\_features() & -0.31 & 0.68\\
\addlinespace[0.3em]
\multicolumn{4}{l}{\textbf{Trend \& Decomposition}}\\
\hspace{1em}STL Trend Strength & stl\_features() & 0.28 & 0.20\\
\hspace{1em}STL Residual ACF (Lag 1) & stl\_features() & 0.71 & -0.28\\
\hspace{1em}STL Residual ACF (Lags 1-10) & stl\_features() & 0.67 & -0.35\\
\hspace{1em}STL Spikiness & stl\_features() & 0.77 & -0.55\\
\hspace{1em}STL Linearity & stl\_features() & -0.91 & 0.28\\
\hspace{1em}STL Curvature & stl\_features() & 0.26 & 0.24\\
\hspace{1em}Holt's Alpha & holt\_parameters() & 0.64 & -0.45\\
\hspace{1em}Holt's Beta & holt\_parameters() & 0.65 & -0.29\\
\addlinespace[0.3em]
\multicolumn{4}{l}{\textbf{Volatility \& Nonlinear Dynamics}}\\
\hspace{1em}ARCH ACF Heterogeneity & heterogeneity() & 0.15 & 0.37\\
\hspace{1em}GARCH ACF Heterogeneity & heterogeneity() & 0.23 & 0.05\\
\hspace{1em}ARCH R² Heterogeneity & heterogeneity() & 0.61 & -0.47\\
\hspace{1em}GARCH R² Heterogeneity & heterogeneity() & 0.49 & -0.51\\
\hspace{1em}ARCH LM Statistic & arch\_stat() & 0.73 & -0.40\\
\hspace{1em}Nonlinearity & nonlinearity() & -0.29 & 0.02\\
\hspace{1em}2D Embedding In-Circle (r=1) & autocorr\_features() & 0.02 & 0.47\\
\hspace{1em}2D Embedding In-Circle (r=2) & autocorr\_features() & 0.03 & 0.47\\
\hspace{1em}Motif Entropy (2-symbol) & autocorr\_features() & -0.54 & -0.37\\
\hspace{1em}Time-Reversibility Statistic & autocorr\_features() & 0.63 & -0.46\\
\hspace{1em}Symbolic RW Crossing Rate & autocorr\_features() & -0.67 & 0.43\\
\hspace{1em}Outlier-Inclusion MDRMD & dist\_features() & -0.92 & -0.03\\
\addlinespace[0.3em]
\multicolumn{4}{l}{\textbf{Distributional Summary}}\\
\hspace{1em}Crossing Points & crossing\_points() & -0.55 & 0.63\\
\hspace{1em}Flat Spots & flat\_spots() & 0.19 & 0.77\\
\hspace{1em}Lumpiness & lumpiness() & 0.31 & -0.30\\
\hspace{1em}Stability & stability() & 0.07 & 0.27\\
\hspace{1em}Proportion of Zeros & zero\_proportion() & -0.07 & 0.57\\
\hspace{1em}Spectral Entropy & entropy() & 0.29 & -0.37\\
\hspace{1em}Hurst Exponent & hurst() & 0.10 & 0.53\\
\hspace{1em}Histogram Mode (10 bins) & dist\_features() & 0.79 & -0.48\\
\addlinespace[0.3em]
\multicolumn{4}{l}{\textbf{Regime Shifts \& Scale}}\\
\hspace{1em}Max Level Shift & max\_level\_shift() & 0.72 & -0.37\\
\hspace{1em}Time of Max Level Shift & max\_level\_shift() & -0.67 & 0.41\\
\hspace{1em}Max Variance Shift & max\_var\_shift() & 0.74 & -0.36\\
\hspace{1em}Time of Max Variance Shift & max\_var\_shift() & -0.68 & 0.43\\
\hspace{1em}Max KL Divergence Shift & max\_kl\_shift() & 0.79 & 0.10\\
\hspace{1em}Time of Max KL Shift & max\_kl\_shift() & -0.72 & 0.41\\
\hspace{1em}Total Case Count & custom & 0.99 & 0.00\\
\bottomrule
\multicolumn{4}{l}{\rule{0pt}{1em}Total number of features: 49}\\
\end{tabular}
\end{table}

\begin{table}[h!]

\caption{\label{tab:treefeatures}Spearman correlation of tree features with parameters}
\centering
\begin{tabular}[t]{llrr}
\toprule
Feature & \texttt{treestats} fn & $R_0$ & $D$\\
\midrule
\addlinespace[0.3em]
\multicolumn{4}{l}{\textbf{Balance \& Topology}}\\
\hspace{1em}Colless Index & colless() & -0.35 & -0.01\\
\hspace{1em}Corrected Colless Index & colless\_corr() & -0.95 & -0.00\\
\hspace{1em}Quadratic Colless Index & colless\_quad() & -0.02 & -0.01\\
\hspace{1em}Equal-Weights Colless & ew\_colless() & -0.95 & -0.00\\
\hspace{1em}Sackin Index & sackin() & -0.05 & -0.01\\
\hspace{1em}B1 Balance Index & b1() & 0.99 & 0.00\\
\hspace{1em}B2 Balance Index & b2() & 0.72 & 0.00\\
\hspace{1em}Beta Statistic & beta\_statistic() & 0.95 & 0.00\\
\hspace{1em}Blum Statistic & blum() & 0.46 & 0.00\\
\hspace{1em}Rogers Index & rogers() & 0.72 & 0.01\\
\hspace{1em}Root Imbalance & root\_imbalance() & -0.15 & -0.01\\
\hspace{1em}Rquartet Index & rquartet() & 0.87 & 0.00\\
\hspace{1em}Number of Cherries & cherries() & 0.99 & 0.00\\
\hspace{1em}Double Cherries & double\_cherries() & 0.96 & 0.00\\
\hspace{1em}Four-Prongs & four\_prong() & 0.80 & 0.01\\
\hspace{1em}Pitchforks & pitchforks() & 0.97 & 0.00\\
\hspace{1em}Stairs & stairs() & -0.88 & -0.00\\
\hspace{1em}Stairs2 & stairs2() & 0.92 & 0.00\\
\hspace{1em}Average Ladder Length & avg\_ladder() & -0.83 & -0.00\\
\hspace{1em}Max Ladder Length & max\_ladder() & -0.43 & 0.00\\
\hspace{1em}IL Number & ILnumber() & 0.40 & 0.00\\
\hspace{1em}Symmetric Nodes & sym\_nodes() & 0.73 & 0.01\\
\hspace{1em}Total Path Length & tot\_path\_length() & -0.09 & -0.01\\
\addlinespace[0.3em]
\multicolumn{4}{l}{\textbf{Branch Length \& Distance}}\\
\hspace{1em}Mean Branch Length & mean\_branch\_length() & 0.02 & 0.99\\
\hspace{1em}Mean Internal Branch Length & mean\_branch\_length\_int() & -0.56 & 0.77\\
\hspace{1em}Mean External Branch Length & mean\_branch\_length\_ext() & 0.34 & 0.92\\
\hspace{1em}Variance of Branch Length & var\_branch\_length() & 0.29 & 0.94\\
\hspace{1em}Internal Branch Length Var. & var\_branch\_length\_int() & -0.46 & 0.81\\
\hspace{1em}External Branch Length Var. & var\_branch\_length\_ext() & 0.39 & 0.89\\
\hspace{1em}Total Cophenetic Index & tot\_coph() & 0.14 & -0.00\\
\hspace{1em}Area per Pair & area\_per\_pair() & -0.57 & -0.01\\
\hspace{1em}Mean Pairwise Distance & mean\_pair\_dist() & -0.50 & 0.67\\
\hspace{1em}Total Internal Path Length & tot\_internal\_path() & -0.12 & -0.01\\
\hspace{1em}Wiener Index & wiener() & -0.03 & 0.68\\
\addlinespace[0.3em]
\multicolumn{4}{l}{\textbf{Tree Shape, Size \& Centrality}}\\
\hspace{1em}Tree Diameter & diameter() & -0.57 & -0.01\\
\hspace{1em}Max Depth & max\_depth() & -0.58 & -0.01\\
\hspace{1em}Max Width & max\_width() & 0.93 & -0.00\\
\hspace{1em}Average Leaf Depth & average\_leaf\_depth() & -0.58 & -0.01\\
\hspace{1em}Max Delta Width & max\_del\_width() & 0.92 & -0.00\\
\hspace{1em}Variance of Depth & var\_leaf\_depth() & -0.66 & -0.00\\
\hspace{1em}Average Vertex Depth & avg\_vert\_depth() & -0.58 & -0.01\\
\hspace{1em}Number of Lineages & number\_of\_lineages() & 0.99 & 0.00\\
\hspace{1em}Treeness & treeness() & -0.97 & -0.00\\
\hspace{1em}Max Width / Max Depth & mw\_over\_md() & 0.94 & 0.00\\
\hspace{1em}Max Betweenness & max\_betweenness() & 0.93 & 0.01\\
\hspace{1em}Max Closeness & max\_closeness(weight=F) & -0.04 & 0.01\\
\hspace{1em}Max Weighted Closeness & max\_closeness(weight=T) & 0.25 & -0.63\\
\addlinespace[0.3em]
\multicolumn{4}{l}{\textbf{Diversification \& Information Content}}\\
\hspace{1em}Fusco-Cronk I Statistic & mean\_i() & -0.94 & -0.00\\
\hspace{1em}Shao \& Sokal J1 Index & j\_one() & 0.76 & 0.01\\
\hspace{1em}Entropy-Based J Statistic & entropy\_j() & -0.81 & 0.50\\
\hspace{1em}Phylo. Species Variability & psv() & -0.50 & 0.67\\
\bottomrule
\multicolumn{4}{l}{\rule{0pt}{1em}Total number of features: 51}\\
\end{tabular}
\end{table}

\subsection{Neural ratio estimation}
\label{subsec:nre}

We used neural ratio estimation (NRE) \cite{durkan_contrastive_2020, hermans2020likelihood} to approximate the joint posterior distribution of $\boldsymbol{\theta}=(R_0,D)^\mathrm{T}$ given outbreak summaries. Below, $\mathbf{x}$ denotes the preprocessed feature vector, including the missing-value mask described in Section \ref{subsec:preprocessing}.

\subsubsection{Inference framework}

NRE estimates the likelihood-to-evidence ratio by training a neural network to distinguish parameter--feature pairs generated together from pairs generated independently. The positive class ($C=1$) consists of pairs from the joint distribution $p(\boldsymbol{\theta},\mathbf{x})$, while the negative class ($C=0$) consists of pairs from the product of marginals $p(\boldsymbol{\theta})p(\mathbf{x})$. With equally weighted classes, the optimal classifier probability $d^*(\boldsymbol{\theta},\mathbf{x})=P(C=1\mid\boldsymbol{\theta},\mathbf{x})$ satisfies
\begin{equation}
    \frac{d^*(\boldsymbol{\theta},\mathbf{x})}
    {1-d^*(\boldsymbol{\theta},\mathbf{x})}
    =
    \frac{p(\boldsymbol{\theta},\mathbf{x})}
    {p(\boldsymbol{\theta})p(\mathbf{x})}
    =
    \frac{p(\mathbf{x}\mid\boldsymbol{\theta})}{p(\mathbf{x})}
    =
    r(\boldsymbol{\theta},\mathbf{x}).
\end{equation}
Thus, the classifier odds recover the likelihood-to-evidence ratio. Bayes' theorem then gives
\begin{equation}
    p(\boldsymbol{\theta}\mid\mathbf{x})
    =
    r(\boldsymbol{\theta},\mathbf{x})p(\boldsymbol{\theta}),
\end{equation}
where $p(\boldsymbol{\theta})$ is the prior distribution underlying the training design.

In practice, the neural classifier approximates $d^*$ with a learned function $\hat d$. For a new observation $\mathbf{x}_{\mathrm{test}}$, the estimated ratio is
\begin{equation}
    \hat r(\boldsymbol{\theta},\mathbf{x}_{\mathrm{test}})
    =
    \frac{\hat d(\boldsymbol{\theta},\mathbf{x}_{\mathrm{test}})}
    {1-\hat d(\boldsymbol{\theta},\mathbf{x}_{\mathrm{test}})}.
\end{equation}
Multiplying this ratio by the prior and normalizing over the parameter space yields an approximate posterior:
\begin{equation}
    \hat p(\boldsymbol{\theta}\mid\mathbf{x}_{\mathrm{test}})
    =
    \frac{
        \hat r(\boldsymbol{\theta},\mathbf{x}_{\mathrm{test}})
        p(\boldsymbol{\theta})
    }{
        \int
        \hat r(\boldsymbol{\theta}',\mathbf{x}_{\mathrm{test}})
        p(\boldsymbol{\theta}')\,d\boldsymbol{\theta}'
    }.
\end{equation}
This procedure enables posterior inference without explicit likelihood evaluation.

\subsubsection{Posterior computation}

\begin{figure}
    \centering
    \includegraphics[width=1\linewidth]{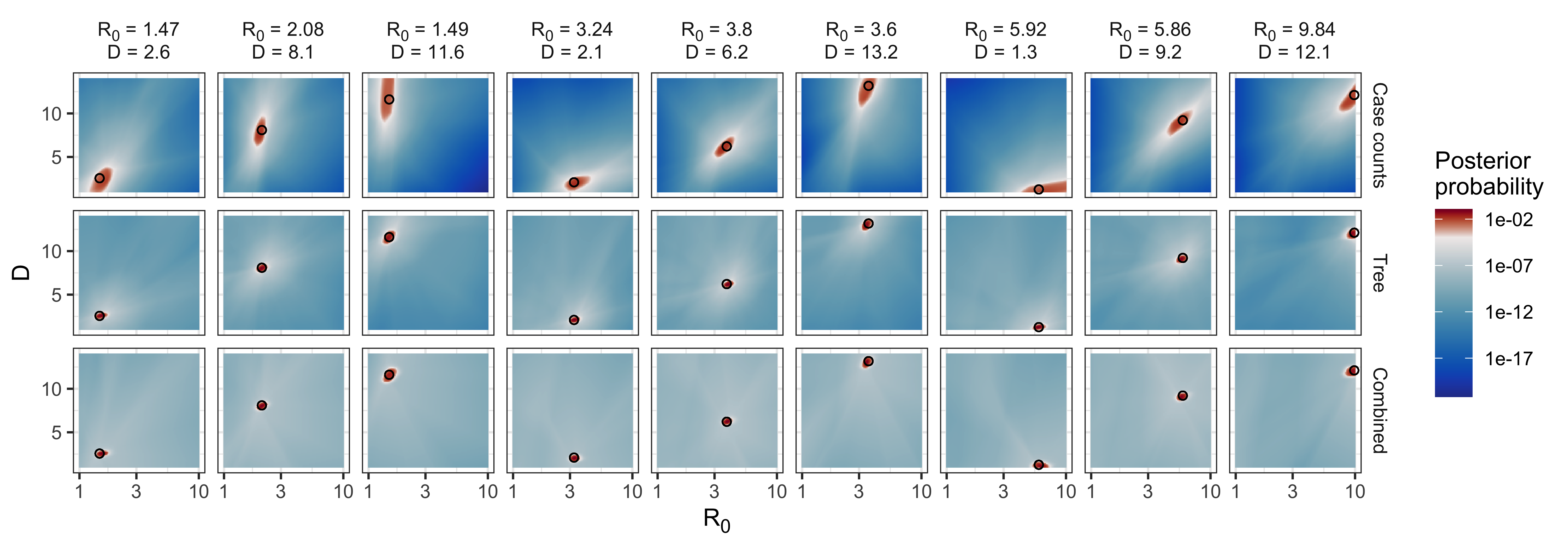}
    \caption{Estimated joint posterior distributions of the basic reproduction number, $R_0$, and average infection duration, $D$, for some test simulations. Columns correspond to outbreaks and rows to NSBI models using case count summaries, transmission tree summaries, or both combined. Colors represent posterior probability mass $\hat w_j$ on a $100\times100$ parameter grid, displayed on a $\log_{10}$ scale. White denotes a cell probability of $10^{-4}$, corresponding to equal probability across all grid cells; blue and red indicate lower and higher probabilities, respectively. Open black circles mark the true parameter values. The nine columns correspond to the nine panels shown in Fig \ref{fig:simulation_examples}A.}
    \label{fig:posterior}
\end{figure}

For each simulation in the test set, we held its features $\mathbf{x}_{\mathrm{test}}$ fixed and evaluated the trained classifier over a $100\times100$ parameter grid spanning the simulation domain. For a grid cell indexed by $j$, posterior probability masses can be approximated as
\begin{equation}
    \hat w_j
    =
    \frac{
        \hat r(\boldsymbol{\theta}_j,\mathbf{x}_{\mathrm{test}})
        p(\boldsymbol{\theta}_j)\Delta_j
    }{
        \sum_k
        \hat r(\boldsymbol{\theta}_k,\mathbf{x}_{\mathrm{test}})
        p(\boldsymbol{\theta}_k)\Delta_k
    },
\end{equation}
where $\sum_{j} \hat{w}_j = 1$, $\boldsymbol{\theta}_j$ is the representative parameter value for cell $j$, and $\Delta_j$ is its area in the coordinates used to define the prior density. In our work, $\Delta_j$ was the same area for all cells $j$. Together, the grid cell-level posterior masses $\hat w_j$ approximate the joint posterior distribution of $(R_0,D)$, representing uncertainty in both parameters and their dependence. Figure \ref{fig:posterior} shows examples of these estimated joint distributions for individual test outbreaks under each of the three NSBI models. The posterior masses $\hat w_j$ were used to calculate the point estimates, marginal intervals, and evaluation metrics described in Section \ref{subsec:evaluation}.

\subsubsection{Implementation and training}

We trained separate neural network classifiers using case count summaries, transmission tree summaries, or their combination. Each classifier was implemented as a feedforward neural network in PyTorch \cite{paszke_pytorch_2019}. The network took the concatenated standardized parameters and preprocessed features as input, and contained three hidden layers with 128, 64, and 32 nodes. Each hidden layer comprised a fully connected linear transformation, batch normalization \cite{ioffe_batch_2015}, a ReLU activation \cite{glorot_deep_2011}, and dropout with probability 0.2 \cite{srivastava_dropout_2014}. The output layer returned a single logit, $\hat h(\boldsymbol{\theta},\mathbf{x})$, without an activation function. The corresponding classifier probability and estimated ratio were
\begin{equation}
    \hat d(\boldsymbol{\theta},\mathbf{x})
    =
    \frac{1}{1+\exp[-\hat h(\boldsymbol{\theta},\mathbf{x})]}.
\end{equation}

For each training batch, positive pairs were taken from the simulated training data, and an equal number of negative pairs were constructed by randomly permuting the parameter vectors to pair features with parameters other than those that generated them. Positive and negative pairs were combined and randomly shuffled within each training batch before being passed through the network. We minimized binary cross-entropy loss with logits using the Adam optimizer \cite{kingma_adam_2017}, with an initial learning rate of $5\times10^{-4}$, weight decay of $10^{-4}$, batch size of 64, and gradient clipping at a maximum norm of 1.0. The \texttt{ReduceLROnPlateau} scheduler reduced the learning rate by a factor of 0.5 when validation loss plateaued (patience of five epochs), down to a minimum of $10^{-6}$. Training stopped when validation loss failed to improve by at least $10^{-4}$ for 100 consecutive epochs or after 10,000 epochs. We retained the checkpoint with the lowest validation loss.

\subsection{Maximum likelihood baseline}
\label{subsec:baseline_methods}

To benchmark the NRE approach, we implemented a classical maximum likelihood estimator based on a deterministic SIR model fit to the case count time series. The model followed the ordinary differential equations
\begin{align}
    \frac{dS}{dt} &= -\frac{\beta SI}{N},\\
    \frac{dI}{dt} &= \frac{\beta SI}{N}-\gamma I,\\
    \frac{dR}{dt} &= \gamma I,
\end{align}
where $N=S+I+R$ was constant and the initial conditions matched those used for the stochastic simulations: $S(0)=4,999$, $I(0)=1$, and $R(0)=0$.

We optimized directly over $R_0$ and $D$ and transformed these parameters to the corresponding transmission and recovery rates within the likelihood calculation according to
\begin{equation}
    \gamma=\frac{1}{D},
    \qquad
    \beta=\frac{R_0}{D}.
\end{equation}
For a candidate parameter pair $(R_0,D)$, the deterministic SIR system was solved numerically using the R package \texttt{deSolve} \cite{desolve}. The expected number of recoveries (cases) during day $t$ was calculated from the change in the cumulative recovered population,
\begin{equation}
    \lambda_t(R_0,D)
    =
    R(t;R_0,D)-R(t-1;R_0,D).
\end{equation}
Observed daily recovery counts $y_t$ were assumed conditionally independent with
\begin{equation}
    y_t \mid R_0,D
    \sim
    \operatorname{Poisson}\!\left(\lambda_t(R_0,D)\right),
\end{equation}
giving the negative log-likelihood
\begin{equation}
    -\ell(R_0,D)
    =
    -\sum_{t=1}^{T}
    \log
    \operatorname{Poisson}
    \!\left(
        y_t \mid \lambda_t(R_0,D)
    \right).
\end{equation}

Maximum likelihood estimates were obtained by minimizing the negative log-likelihood using the L-BFGS-B algorithm \cite{byrd1995}, as implemented in the R package \texttt{bbmle} \cite{bolker2008}. Optimization bounds were set 10\% wider than the parameter ranges used for simulation. Starting values were set to the geometric midpoint of the $R_0$ range and the arithmetic midpoint of the $D$ range.

Approximate 95\% confidence intervals were constructed from the asymptotic multivariate normal approximation to the sampling distribution of the maximum likelihood estimator. The covariance matrix was obtained from the inverse observed information matrix evaluated at the maximum likelihood estimate. We drew 10,000 parameter vectors from this multivariate normal distribution using the R package \texttt{mvtnorm} \cite{genz_computation_2009} and defined the marginal 95\% confidence interval for each parameter using the 2.5th and 97.5th percentiles of the corresponding draws. Point estimates and confidence intervals were evaluated using the criteria described in Section \ref{subsec:evaluation}.

\subsection{Evaluation}
\label{subsec:evaluation}

We evaluated parameter estimation using point-estimate accuracy, 95\% uncertainty interval widths, 95\% empirical coverages, and the interval score. Metrics were calculated separately for $R_0$ and $D$. For the NRE models, posterior-based quantities were calculated from the discrete posterior masses $\hat w_j$ described in Section \ref{subsec:nre}. The baseline was evaluated using its maximum likelihood estimates and corresponding 95\% confidence intervals.

\subsubsection{Point-estimate accuracy}

For scalar parameter $\theta\in\{R_0,D\}$ and test simulation $i$, absolute error was defined as
\begin{equation}
    \operatorname{AE}_{i,\theta}
    =
    \left|
        \hat{\theta}_i-\theta_i
    \right|,
\end{equation}
where $\theta_i$ is the true parameter value and $\hat{\theta}_i$ is its point estimate. For the NRE models, $\hat{\theta}_i$ was the posterior mean obtained from the corresponding marginal posterior distribution. For the baseline, $\hat{\theta}_i$ was the maximum likelihood estimate. Mean absolute error (MAE) across the $n_{\mathrm{test}} = 20,000$ test simulations was
\begin{equation}
    \operatorname{MAE}_{\theta}
    =
    \frac{1}{n_{\mathrm{test}}}
    \sum_{i=1}^{n_{\mathrm{test}}}
    \operatorname{AE}_{i,\theta}.
\end{equation}
Lower MAE indicates more accurate point estimation.

\subsubsection{Uncertainty interval size}

For each parameter separately, we calculated the width of its 95\% uncertainty interval as
\begin{equation}
    W_{i,\theta}
    =
    U_{i,\theta}-L_{i,\theta},
\end{equation}
where $L_{i,\theta}$ and $U_{i,\theta}$ denote the lower and upper bounds of the interval. For the NRE models, marginal posterior distributions for $R_0$ and $D$ were obtained by summing the joint posterior masses over the other parameter, and 95\% credible intervals were constructed from these marginal distributions. For the baseline, the 95\% confidence intervals described in Section \ref{subsec:baseline_methods} were used. Smaller interval widths indicate greater precision, although width must be considered jointly with coverage.

\subsubsection{Empirical coverage}

Marginal 95\% interval coverage for parameter $\theta$ was recorded for each test simulation as
\begin{equation}
    C_{i,\theta}
    =
    \mathbf{I}
    (
        L_{i,\theta}
        \leq
        \theta_i
        \leq
        U_{i,\theta}
    ),
\end{equation}
where $\mathbf{I}(\cdot)$ is the indicator function equal to 1 if the argument is true and 0 otherwise. Empirical coverage is then calculate as
\begin{equation}
    \widehat{\operatorname{Cov}}_{\theta}
    =
    \frac{1}{n_{\mathrm{test}}}
    \sum_{i=1}^{n_{\mathrm{test}}}
    C_{i,\theta}.
\end{equation}
A well-calibrated nominal 95\% interval procedure is expected to achieve empirical coverage near 0.95 over repeated test simulations.

\subsubsection{Interval score}

To evaluate interval sharpness and calibration jointly, we calculated the interval score \cite{winkler_decision-theoretic_1972} for each marginal 95\% interval. For a central coverage level of $1-\alpha$, with lower and upper bounds $L$ and $U$ and true parameter value $\theta$, the interval score is
\begin{equation}
    \operatorname{IS}_{\alpha}(L,U;\theta)
    =
    (U-L)
    +
    \frac{2}{\alpha}(L-\theta)
    \mathbf{I}(\theta<L)
    +
    \frac{2}{\alpha}(\theta-U)
    \mathbf{I}(\theta>U).
\end{equation}
We used $\alpha=0.05$ for the 95\% intervals. The first term rewards narrow intervals, while the remaining terms penalize intervals that fail to contain the true parameter value, with larger penalties for larger misses. Lower interval scores indicate better performance.

\subsection{Software and packages}
\label{subsec:software}

Snakemake v9.24.0 \cite{molder_sustainable_2021} was used for workflow orchestration. Data processing was done in Python v3.13.15 \cite{python3} with the packages NumPy v2.2.6 \cite{harris2020array}, and pandas v2.3.3 \cite{reback2020pandas, mckinney-proc-scipy-2010}. R v4.4.3 \cite{r_core_team_r_nodate} was used for all data analysis and visualization with the packages tidyverse v2.0.0 \cite{wickham_welcome_2019}, scales v1.4.0 \cite{wickham_scales_2025}, ggpubr v0.6.2 \cite{kassambara_ggpubr_2026}, ggtree v4.1.1.6 \cite{yu_ggtree_2017}, ggtext v0.1.2 \cite{ggtext}, patchwork v1.3.2 \cite{patchwork}, kableExtra v1.4.0 \cite{kableextra}, and cmocean v0.3.2 \cite{thyng2016cmocean}.

\section{Data and code availability}

All code, configuration files, and analysis scripts are available at https://github.com/lanl/precog/tree/main/nsbi.

\section{Acknowledgments}
We are grateful to Nate Kornetzke for helpful discussions and feedback.
The authors acknowledge the use of LLMs for text editing and assistance with selected coding tasks.
The authors retain full responsibility for the manuscript's content.
This article has been approved for unlimited release under LA-UR-26-28863.

\section{Funding}
ZL, EG, JT, TS, DO, and LC were supported by the Laboratory Directed Research and Development program of Los Alamos National Laboratory under project number 20240066DR. 

\section{Author contributions}

Conceptualization: ZL, EG, DO, LC\\
Formal analysis: ZL\\
Funding acquisition: LC, DO, EG\\
Investigation: ZL\\
Methodology: ZL, EG, JT, TS, DO\\
Project administration: LC, DO\\
Software: ZL\\
Supervision: LC, DO\\
Visualization: ZL\\
Writing - original draft: ZL, DO\\
Writing - review \& editing: ZL, EG, JT, TS, DO

\bibliographystyle{vancouver}
\bibliography{references}

@article{cuartero_impact_2026,
	title = {The impact of ambiguously reported epidemiological parameters for infectious disease modelling and recommended best practices},
	issn = {1755-4365},
	url = {https://www.sciencedirect.com/science/article/pii/S1755436526000587},
	doi = {10.1016/j.epidem.2026.100942},
	urldate = {2026-09-21},
	journal = {Epidemics},
	author = {Cuartero, Carmen Tamayo and Lambert, Joshua W. and Cuomo-Dannenburg, Gina and Fitzner, Julia and Ndumbi, Patricia and Rice, Chloe and Saulo, Dina and Waddell, Lisa and Azam, James M. and Doohan, Patrick and Hicks, Joseph T. and McCain, Kelly and Morgenstern, Christian and Naidoo, Tristan M. and Ruybal-Pesántez, Shazia and Rawson, Thomas and Santoni, Cosmo and Unwin, H. Juliette T. and Kucharski, Adam and Cori, Anne and Bhatia, Sangeeta and McCabe, Ruth},
	month = aug,
	year = {2026},
	pages = {100942},
}

@article{ridenhour_unraveling_2014,
	title = {Unraveling {R0}: {Considerations} for {Public} {Health} {Applications}},
	volume = {104},
	issn = {0090-0036},
	shorttitle = {Unraveling {R0}},
	url = {https://pmc.ncbi.nlm.nih.gov/articles/PMC3935673/},
	doi = {10.2105/AJPH.2013.301704},
	number = {2},
	urldate = {2026-09-21},
	journal = {American Journal of Public Health},
	author = {Ridenhour, Benjamin and Kowalik, Jessica M. and Shay, David K.},
	month = feb,
	year = {2014},
	pages = {e32--e41},
}

@article{boonpatcharanon_estimating_2022,
	title = {Estimating the basic reproduction number at the beginning of an outbreak},
	volume = {17},
	issn = {1932-6203},
	url = {https://journals.plos.org/plosone/article?id=10.1371/journal.pone.0269306},
	doi = {10.1371/journal.pone.0269306},
	language = {en},
	number = {6},
	urldate = {2026-09-21},
	journal = {PLOS ONE},
	publisher = {Public Library of Science},
	author = {Boonpatcharanon, Sawitree and Heffernan, Jane M. and Jankowski, Hanna},
	month = jun,
	year = {2022},
	pages = {e0269306},
}

@article{cranmer_frontier_2020,
	title = {The frontier of simulation-based inference},
	volume = {117},
	url = {https://www.pnas.org/doi/10.1073/pnas.1912789117},
	doi = {10.1073/pnas.1912789117},
	number = {48},
	urldate = {2026-09-21},
	journal = {Proceedings of the National Academy of Sciences},
	publisher = {Proceedings of the National Academy of Sciences},
	author = {Cranmer, Kyle and Brehmer, Johann and Louppe, Gilles},
	month = dec,
	year = {2020},
	pages = {30055--30062},
}

@article{zammit-mangion_neural_2025,
	title = {Neural {Methods} for {Amortized} {Inference}},
	volume = {12},
	issn = {2326-831X},
	url = {https://www.annualreviews.org/content/journals/10.1146/annurev-statistics-112723-034123},
	doi = {10.1146/annurev-statistics-112723-034123},
	urldate = {2026-09-21},
	journal = {Annual Review of Statistics and Its Application},
	publisher = {Annual Reviews},
	author = {Zammit-Mangion, Andrew and Sainsbury-Dale, Matthew and Huser, Raphaël},
	year = {2025},
	pages = {311--335},
}

@article{jang_comparative_2026,
	title = {A comparative study of simulation-based inference methods for epidemic models with identifiability considerations},
	volume = {22},
	issn = {1553-7358},
	url = {https://journals.plos.org/ploscompbiol/article?id=10.1371/journal.pcbi.1014364},
	doi = {10.1371/journal.pcbi.1014364},
	language = {en},
	number = {6},
	urldate = {2026-09-21},
	journal = {PLOS Computational Biology},
	publisher = {Public Library of Science},
	author = {Jang, Geunsoo and Candan, K. Selçuk and Chowell, Gerardo},
	month = jun,
	year = {2026},
	pages = {e1014364},
}

@article{radev_outbreakflow_2021,
	title = {{OutbreakFlow}: {Model}-based {Bayesian} inference of disease outbreak dynamics with invertible neural networks and its application to the {COVID}-19 pandemics in {Germany}},
	volume = {17},
	issn = {1553-734X},
	shorttitle = {{OutbreakFlow}},
	url = {https://pmc.ncbi.nlm.nih.gov/articles/PMC8584772/},
	doi = {10.1371/journal.pcbi.1009472},
	number = {10},
	urldate = {2026-09-21},
	journal = {PLoS Computational Biology},
	author = {Radev, Stefan T. and Graw, Frederik and Chen, Simiao and Mutters, Nico T. and Eichel, Vanessa M. and Bärnighausen, Till and Köthe, Ullrich},
	month = oct,
	year = {2021},
	pages = {e1009472},
}

@misc{chatha_neural_2026,
	title = {Neural {Posterior} {Estimation} for {Stochastic} {Epidemic} {Modeling}},
	url = {http://arxiv.org/abs/2412.12967},
	doi = {10.48550/arXiv.2412.12967},
	language = {en},
	urldate = {2026-09-21},
	publisher = {arXiv},
	author = {Chatha, Prayag and Bu, Fan and Regier, Jeffrey and Snitkin, Evan and Zelner, Jon},
	month = jun,
	year = {2026},
	note = {arXiv:2412.12967 [stat.ME]},
}

@article{garot2026teddy,
  title={Teddy: neural inference of epidemiological parameters from viral sequences},
  author={Garot, Vincent and Blassel, Luc and Nesterenko, Luca and Zhukova, Anna and Alizon, Samuel and Jacob, Laurent},
  journal={bioRxiv},
  pages={2026--01},
  year={2026},
  publisher={Cold Spring Harbor Laboratory}
}

@misc{zarebski_amortized_2026,
	title = {Amortized {Phylodynamic} {Inference} with {Neural} {Bayes} {Estimators} and {Recursive} {Neural} {Networks}},
	url = {http://arxiv.org/abs/2603.08345},
	doi = {10.48550/arXiv.2603.08345},
	language = {en},
	urldate = {2026-09-21},
	publisher = {arXiv},
	author = {Zarebski, Alexander E. and Williams, Thomas and Plessis, Louis du},
	month = mar,
	year = {2026},
	note = {arXiv:2603.08345 [stat.ME]},
}

@article{pinotti2026simulation,
  title={Simulation-based inference of epidemiological and phylodynamic models via neural posterior estimation},
  author={Pinotti, Francesco and Th{\'e}z{\'e}, Julien and Bailly, Xavier and Fourni{\'e}, Guillaume},
  journal={Proceedings of the Royal Society B: Biological Sciences},
  volume={293},
  number={2077},
  year={2026},
  publisher={The Royal Society}
}

@article{perez_phylocnn_2026,
	title = {{PhyloCNN}: {Improving} {Tree} {Representation} and {Neural} {Network} {Architecture} for {Deep} {Learning} from {Trees} in {Phylodynamics} and {Diversification} {Studies}},
	volume = {75},
	issn = {1063-5157},
	shorttitle = {{PhyloCNN}},
	url = {https://doi.org/10.1093/sysbio/syaf082},
	doi = {10.1093/sysbio/syaf082},
	number = {4},
	urldate = {2026-09-21},
	journal = {Systematic Biology},
	author = {Perez, Manolo Fernandez and Gascuel, Olivier},
	month = jul,
	year = {2026},
	pages = {776--795},
}

@article{judge_epifusion_2024,
	title = {{EpiFusion}: {Joint} inference of the effective reproduction number by integrating phylodynamic and epidemiological modelling with particle filtering},
	volume = {20},
	issn = {1553-7358},
	shorttitle = {{EpiFusion}},
	url = {https://journals.plos.org/ploscompbiol/article?id=10.1371/journal.pcbi.1012528},
	doi = {10.1371/journal.pcbi.1012528},
	language = {en},
	number = {11},
	urldate = {2026-07-29},
	journal = {PLOS Computational Biology},
	publisher = {Public Library of Science},
	author = {Judge, Ciara and Vaughan, Timothy and Russell, Timothy and Abbott, Sam and Plessis, Louis du and Stadler, Tanja and Brady, Oliver and Hill, Sarah},
	month = nov,
	year = {2024},
	pages = {e1012528},
}

@inproceedings{durkan_contrastive_2020,
	title = {On {Contrastive} {Learning} for {Likelihood}-free {Inference}},
	issn = {2640-3498},
	url = {https://proceedings.mlr.press/v119/durkan20a.html},
	language = {en},
	urldate = {2026-09-21},
	booktitle = {Proceedings of the 37th {International} {Conference} on {Machine} {Learning}},
	publisher = {PMLR},
	author = {Durkan, Conor and Murray, Iain and Papamakarios, George},
	month = nov,
	year = {2020},
	pages = {2771--2781},
}

@inproceedings{hermans2020likelihood,
  title={Likelihood-free MCMC with amortized approximate ratio estimators},
  author={Hermans, Joeri and Begy, Volodimir and Louppe, Gilles},
  booktitle={International conference on machine learning},
  pages={4239--4248},
  year={2020},
  organization={PMLR}
}

@misc{noauthor_clicumupydoe2_2026,
	title = {clicumu/{pyDOE2}},
	copyright = {BSD-3-Clause},
	url = {https://github.com/clicumu/pyDOE2},
	urldate = {2026-09-21},
	publisher = {clicumu},
	month = sep,
	year = {2026},
	note = {original-date: 2016-09-22T05:05:21Z},
}

@article{mckay_comparison_1979,
	title = {A {Comparison} of {Three} {Methods} for {Selecting} {Values} of {Input} {Variables} in the {Analysis} of {Output} from a {Computer} {Code}},
	volume = {21},
	issn = {0040-1706},
	url = {https://www.jstor.org/stable/1268522},
	doi = {10.2307/1268522},
	number = {2},
	urldate = {2026-09-22},
	journal = {Technometrics},
	publisher = {[Taylor \& Francis, Ltd., American Statistical Association, American Society for Quality]},
	author = {McKay, M. D. and Beckman, R. J. and Conover, W. J.},
	year = {1979},
	pages = {239--245},
}

@article{johnson_minimax_1990,
	title = {Minimax and maximin distance designs},
	volume = {26},
	issn = {0378-3758},
	url = {https://www.sciencedirect.com/science/article/pii/037837589090122B},
	doi = {10.1016/0378-3758(90)90122-B},
	number = {2},
	urldate = {2026-09-22},
	journal = {Journal of Statistical Planning and Inference},
	author = {Johnson, M. E. and Moore, L. M. and Ylvisaker, D.},
	month = oct,
	year = {1990},
	pages = {131--148},
}

@misc{molder_sustainable_2021,
	title = {Sustainable data analysis with {Snakemake}},
	copyright = {https://creativecommons.org/licenses/by/4.0/},
	url = {https://f1000research.com/articles/10-33},
	doi = {10.12688/f1000research.29032.2},
	language = {en},
	urldate = {2025-05-01},
	publisher = {F1000Research},
	author = {Mölder, Felix and Jablonski, Kim Philipp and Letcher, Brice and Hall, Michael B. and Tomkins-Tinch, Christopher H. and Sochat, Vanessa and Forster, Jan and Lee, Soohyun and Twardziok, Sven O. and Kanitz, Alexander and Wilm, Andreas and Holtgrewe, Manuel and Rahmann, Sven and Nahnsen, Sven and Köster, Johannes},
	month = apr,
	year = {2021},
}

@software{python3,
  author = {{Python Software Foundation}},
  title = {Python Language Reference},
  url = {https://www.python.org/},
  year = {2026}
}

@software{reback2020pandas,
    author       = {The pandas development team},
    title        = {pandas-dev/pandas: Pandas},
    month        = feb,
    year         = 2020,
    publisher    = {Zenodo},
    doi          = {10.5281/zenodo.3509134},
    url          = {https://doi.org/10.5281/zenodo.3509134}
}

@InProceedings{ mckinney-proc-scipy-2010,
  author    = { {W}es {M}c{K}inney },
  title     = { {D}ata {S}tructures for {S}tatistical {C}omputing in {P}ython },
  booktitle = { {P}roceedings of the 9th {P}ython in {S}cience {C}onference },
  pages     = { 56 - 61 },
  year      = { 2010 },
  editor    = { {S}t\'efan van der {W}alt and {J}arrod {M}illman },
  doi       = { 10.25080/Majora-92bf1922-00a }
}

@article{wickham_welcome_2019,
	title = {Welcome to the {Tidyverse}},
	volume = {4},
	issn = {2475-9066},
	url = {https://joss.theoj.org/papers/10.21105/joss.01686},
	doi = {10.21105/joss.01686},
	language = {en},
	number = {43},
	urldate = {2024-11-13},
	journal = {Journal of Open Source Software},
	author = {Wickham, Hadley and Averick, Mara and Bryan, Jennifer and Chang, Winston and McGowan, Lucy D'Agostino and François, Romain and Grolemund, Garrett and Hayes, Alex and Henry, Lionel and Hester, Jim and Kuhn, Max and Pedersen, Thomas Lin and Miller, Evan and Bache, Stephan Milton and Müller, Kirill and Ooms, Jeroen and Robinson, David and Seidel, Dana Paige and Spinu, Vitalie and Takahashi, Kohske and Vaughan, Davis and Wilke, Claus and Woo, Kara and Yutani, Hiroaki},
	month = nov,
	year = {2019},
	pages = {1686},
}

@misc{wickham_scales_2025,
	title = {scales: {Scale} {Functions} for {Visualization}},
	copyright = {MIT + file LICENSE},
	shorttitle = {scales},
	url = {https://cran.r-project.org/web/packages/scales/index.html},
	urldate = {2026-03-26},
	author = {Wickham, Hadley and Pedersen, Thomas Lin and Seidel, Dana and Software, Posit and PBC [cph and fnd},
	month = apr,
	year = {2025},
}

@misc{kassambara_ggpubr_2026,
	title = {ggpubr: 'ggplot2' {Based} {Publication} {Ready} {Plots}},
	copyright = {GPL-2 {\textbar} GPL-3 [expanded from: GPL (≥ 2)]},
	shorttitle = {ggpubr},
	url = {https://cran.r-project.org/web/packages/ggpubr/index.html},
	urldate = {2026-03-26},
	author = {Kassambara, Alboukadel and Business, Laszlo Erdey (Faculty of Economics and and Debrecen, University of and Hungary)},
	month = feb,
	year = {2026},
}

@article{paradis_ape_2019,
	title = {ape 5.0: an environment for modern phylogenetics and evolutionary analyses in {R}},
	volume = {35},
	issn = {1367-4803},
	shorttitle = {ape 5.0},
	url = {https://doi.org/10.1093/bioinformatics/bty633},
	doi = {10.1093/bioinformatics/bty633},
	number = {3},
	urldate = {2024-12-06},
	journal = {Bioinformatics},
	author = {Paradis, Emmanuel and Schliep, Klaus},
	month = feb,
	year = {2019},
	pages = {526--528},
}

@article{yu_ggtree_2017,
	title = {ggtree: an r package for visualization and annotation of phylogenetic trees with their covariates and other associated data},
	volume = {8},
	copyright = {© 2016 The Authors. Methods in Ecology and Evolution © 2016 British Ecological Society},
	issn = {2041-210X},
	shorttitle = {ggtree},
	url = {https://onlinelibrary.wiley.com/doi/abs/10.1111/2041-210X.12628},
	doi = {10.1111/2041-210X.12628},
	language = {en},
	number = {1},
	urldate = {2024-12-06},
	journal = {Methods in Ecology and Evolution},
	author = {Yu, Guangchuang and Smith, David K. and Zhu, Huachen and Guan, Yi and Lam, Tommy Tsan-Yuk},
	year = {2017},
	note = {\_eprint: https://onlinelibrary.wiley.com/doi/pdf/10.1111/2041-210X.12628},
	pages = {28--36},
}

@Article{         harris2020array,
 title         = {Array programming with {NumPy}},
 author        = {Charles R. Harris and K. Jarrod Millman and St{\'{e}}fan J.
                 van der Walt and Ralf Gommers and Pauli Virtanen and David
                 Cournapeau and Eric Wieser and Julian Taylor and Sebastian
                 Berg and Nathaniel J. Smith and Robert Kern and Matti Picus
                 and Stephan Hoyer and Marten H. van Kerkwijk and Matthew
                 Brett and Allan Haldane and Jaime Fern{\'{a}}ndez del
                 R{\'{i}}o and Mark Wiebe and Pearu Peterson and Pierre
                 G{\'{e}}rard-Marchant and Kevin Sheppard and Tyler Reddy and
                 Warren Weckesser and Hameer Abbasi and Christoph Gohlke and
                 Travis E. Oliphant},
 year          = {2020},
 month         = sep,
 journal       = {Nature},
 volume        = {585},
 number        = {7825},
 pages         = {357--362},
 doi           = {10.1038/s41586-020-2649-2},
 publisher     = {Springer Science and Business Media {LLC}},
 url           = {https://doi.org/10.1038/s41586-020-2649-2}
}

@misc{r_core_team_r_nodate,
	title = {R: {The} {R} {Project} for {Statistical} {Computing}},
	url = {https://www.r-project.org/},
	urldate = {2024-11-13},
	author = {{R Core Team}},
}

@inproceedings{ioffe_batch_2015,
	title = {Batch {Normalization}: {Accelerating} {Deep} {Network} {Training} by {Reducing} {Internal} {Covariate} {Shift}},
	issn = {1938-7228},
	shorttitle = {Batch {Normalization}},
	url = {https://proceedings.mlr.press/v37/ioffe15.html},
	language = {en},
	urldate = {2026-09-23},
	booktitle = {Proceedings of the 32nd {International} {Conference} on {Machine} {Learning}},
	publisher = {PMLR},
	author = {Ioffe, Sergey and Szegedy, Christian},
	month = jun,
	year = {2015},
	pages = {448--456},
}

@article{srivastava_dropout_2014,
	title = {Dropout: {A} {Simple} {Way} to {Prevent} {Neural} {Networks} from {Overfitting}},
	volume = {15},
	issn = {1533-7928},
	shorttitle = {Dropout},
	url = {http://jmlr.org/papers/v15/srivastava14a.html},
	number = {56},
	urldate = {2026-09-23},
	journal = {Journal of Machine Learning Research},
	author = {Srivastava, Nitish and Hinton, Geoffrey and Krizhevsky, Alex and Sutskever, Ilya and Salakhutdinov, Ruslan},
	year = {2014},
	pages = {1929--1958},
}

@inproceedings{glorot_deep_2011,
	title = {Deep {Sparse} {Rectifier} {Neural} {Networks}},
	issn = {1938-7228},
	url = {https://proceedings.mlr.press/v15/glorot11a.html},
	language = {en},
	urldate = {2026-09-23},
	booktitle = {Proceedings of the {Fourteenth} {International} {Conference} on {Artificial} {Intelligence} and {Statistics}},
	publisher = {JMLR Workshop and Conference Proceedings},
	author = {Glorot, Xavier and Bordes, Antoine and Bengio, Yoshua},
	month = jun,
	year = {2011},
	pages = {315--323},
}

@misc{kingma_adam_2017,
	title = {Adam: {A} {Method} for {Stochastic} {Optimization}},
	shorttitle = {Adam},
	url = {http://arxiv.org/abs/1412.6980},
	doi = {10.48550/arXiv.1412.6980},
	urldate = {2026-09-23},
	publisher = {arXiv},
	author = {Kingma, Diederik P. and Ba, Jimmy},
	month = jan,
	year = {2017},
	note = {arXiv:1412.6980 [cs.LG]},
}

@incollection{paszke_pytorch_2019,
	address = {Red Hook, NY, USA},
	title = {{PyTorch}: an imperative style, high-performance deep learning library},
	shorttitle = {{PyTorch}},
	url = {https://dl.acm.org/doi/10.5555/3454287.3455008},
	number = {721},
	urldate = {2026-09-22},
	booktitle = {Proceedings of the 33rd {International} {Conference} on {Neural} {Information} {Processing} {Systems}},
	publisher = {Curran Associates Inc.},
	author = {Paszke, Adam and Gross, Sam and Massa, Francisco and Lerer, Adam and Bradbury, James and Chanan, Gregory and Killeen, Trevor and Lin, Zeming and Gimelshein, Natalia and Antiga, Luca and Desmaison, Alban and Köpf, Andreas and Yang, Edward and DeVito, Zach and Raison, Martin and Tejani, Alykhan and Chilamkurthy, Sasank and Steiner, Benoit and Fang, Lu and Bai, Junjie and Chintala, Soumith},
	month = dec,
	year = {2019},
	pages = {8026--8037},
}

@article{winkler_decision-theoretic_1972,
	title = {A {Decision}-{Theoretic} {Approach} to {Interval} {Estimation}},
	volume = {67},
	issn = {0162-1459},
	url = {https://www.tandfonline.com/doi/abs/10.1080/01621459.1972.10481224},
	doi = {10.1080/01621459.1972.10481224},
	number = {337},
	urldate = {2026-09-24},
	journal = {Journal of the American Statistical Association},
	publisher = {Taylor \& Francis},
	author = {Winkler, Robert L.},
	month = mar,
	year = {1972},
	pages = {187--191},
}

@book{genz_computation_2009,
	address = {Berlin, Heidelberg},
	series = {Lecture {Notes} in {Statistics}},
	title = {Computation of {Multivariate} {Normal} and t {Probabilities}},
	volume = {195},
	copyright = {https://www.springernature.com/gp/researchers/text-and-data-mining},
	isbn = {978-3-642-01688-2 978-3-642-01689-9},
	url = {https://link.springer.com/10.1007/978-3-642-01689-9},
	doi = {10.1007/978-3-642-01689-9},
	language = {en},
	urldate = {2026-09-24},
	publisher = {Springer},
	author = {Genz, Alan and Bretz, Frank},
	year = {2009},
}

@article{vaughan2024,
  title={{ReMASTER}: improved phylodynamic simulation for {BEAST} 2.7},
  author={Vaughan, Timothy G},
  journal={Bioinformatics},
  volume={40},
  number={1},
  pages={btae015},
  year={2024},
  publisher={Oxford University Press}
}

@article{bouckaert2019,
  title={{BEAST} 2.5: An advanced software platform for {B}ayesian evolutionary analysis},
  author={Bouckaert, Remco and Vaughan, Timothy G and Barido-Sottani, Jo{\"e}lle and Duch{\^e}ne, Sebasti{\'a}n and Fourment, Mathieu and Gavryushkina, Alexandra and Heled, Joseph and Jones, Graham and K{\"u}hnert, Denise and De Maio, Nicola and others},
  journal={PLoS Computational Biology},
  volume={15},
  number={4},
  pages={e1006650},
  year={2019},
  publisher={Public Library of Science San Francisco, CA USA}
}

@Manual{tsfeatures,
  title = {tsfeatures: Time Series Feature Extraction},
  author = {Rob Hyndman and Yanfei Kang and Pablo Montero-Manso and Mitchell O'Hara-Wild and Thiyanga Talagala and Earo Wang and Yangzhuoran Yang},
  year = {2026},
  note = {R package version 1.2.0.0},
  url = {https://pkg.robjhyndman.com/tsfeatures/},
}

@Article{ape,
  title = {ape 5.0: an environment for modern phylogenetics and evolutionary analyses in {R}},
  author = {Emmanuel Paradis and Klaus Schliep},
  journal = {Bioinformatics},
  year = {2019},
  volume = {35},
  pages = {526-528},
  doi = {10.1093/bioinformatics/bty633},
}

@Article{treestats,
  title = {Phylogenetic tree statistics: a systematic overview using the new {R} package `treestats'.},
  author = {Thijs Janzen and Etienne, Rampal S.},
  journal = {Molecular Phylogenetics and Evolution},
  volume = {200},
  pages = {108168},
  year = {2024},
  doi = {10.1016/j.ympev.2024.108168}
}

@article{beaumont2002approximate,
  title={Approximate {B}ayesian Computation in Population Genetics},
  author={Beaumont, Mark A and Zhang, Wenyang and Balding, David J},
  journal={Genetics},
  volume={162},
  number={4},
  pages={2025--2035},
  year={2002},
  publisher={Oxford University Press}
}

@article{byrd1995,
  author    = {Byrd, Richard H. and Lu, Peihuang and Nocedal, Jorge and Zhu, Ciyou},
  title     = {A Limited Memory Algorithm for Bound Constrained Optimization},
  journal   = {{SIAM} Journal on Scientific and Statistical Computing},
  volume    = {16},
  number    = {5},
  pages     = {1190--1208},
  year      = {1995},
  publisher = {SIAM}
}

@book{bolker2008,
  title = {Ecological Models and Data in R},
  author = {Bolker, Benjamin M.},
  year = {2008},
  publisher = {Princeton University Press},
  address = {Princeton, NJ}
}

@Article{desolve,
  title = {Solving Differential Equations in {R}: Package {deSolve}},
  author = {Karline Soetaert and Thomas Petzoldt and R. Woodrow Setzer},
  journal = {Journal of Statistical Software},
  year = {2010},
  volume = {33},
  number = {9},
  pages = {1--25},
  doi = {10.18637/jss.v033.i09},
}

@article{rasmussen2011inference,
  title={Inference for nonlinear epidemiological models using genealogies and time series},
  author={Rasmussen, David A and Ratmann, Oliver and Koelle, Katia},
  journal={{PLoS Computational Biology}},
  volume={7},
  number={8},
  pages={e1002136},
  year={2011},
  publisher={Public Library of Science San Francisco, USA}
}

@article{vaughan2019estimating,
  title={Estimating epidemic incidence and prevalence from genomic data},
  author={Vaughan, Timothy G and Leventhal, Gabriel E and Rasmussen, David A and Drummond, Alexei J and Welch, David and Stadler, Tanja},
  journal={{Molecular Biology and Evolution}},
  volume={36},
  number={8},
  pages={1804--1816},
  year={2019},
  publisher={Oxford University Press}
}

@article{zarebski2022computationally,
  title={A computationally tractable birth-death model that combines phylogenetic and epidemiological data},
  author={Zarebski, Alexander Eugene and du Plessis, Louis and Parag, Kris Varun and Pybus, Oliver George},
  journal={{PLOS Computational Biology}},
  volume={18},
  number={2},
  pages={e1009805},
  year={2022},
  publisher={Public Library of Science San Francisco, CA USA}
}

@Manual{ggtext,
    title = {ggtext: Improved Text Rendering Support for 'ggplot2'},
    author = {Claus O. Wilke and Brenton M. Wiernik},
    year = {2022},
    note = {R package version 0.1.2},
    url = {https://CRAN.R-project.org/package=ggtext},
  }

@Manual{patchwork,
    title = {patchwork: The Composer of Plots},
    author = {Thomas Lin Pedersen},
    year = {2025},
    note = {R package version 1.3.2},
    url = {https://CRAN.R-project.org/package=patchwork},
  }

@Manual{kableextra,
    title = {kableExtra: Construct Complex Table with 'kable' and Pipe Syntax},
    author = {Hao Zhu},
    year = {2024},
    note = {R package version 1.4.0},
    url = {https://CRAN.R-project.org/package=kableExtra},
  }

@Article{thyng2016cmocean,
    title = {True colors of oceanography: Guidelines for effective and accurate colormap selection},
    author = {Kristen M. Thyng and Chad A. Greene and Robert D. Hetland and Heather M. Zimmerle and Steven F. DiMarco},
    doi = {10.5670/oceanog.2016.66},
    journal = {Oceanography},
    number = {3},
    volume = {29},
    year = {2016},
  }
\clearpage





\end{document}